%% file: Portrait_ST.tex
\documentclass[acmtog]{acmart}
\newcommand{\final}{1}
\renewcommand\footnotetextcopyrightpermission[1]{} 
\usepackage{algorithm}
\usepackage{algorithmic}
\usepackage{graphicx}
\usepackage{subfigure}
\usepackage{color}
\usepackage{multirow}
\usepackage{ulem}
\input{macros}

\AtBeginDocument{%
  \providecommand\BibTeX{{%
    \normalfont B\kern-0.5em{\scshape i\kern-0.25em b}\kern-0.8em\TeX}}}

\begin{document}

\title{Towards Harmonized Regional Style Transfer and Manipulation for Facial Images}

\input{Sections/authors.tex}
\input{Sections/0_abstract.tex}
\maketitle
\input{Sections/1_introduction.tex}
\input{Sections/2_related.tex}
\input{Sections/3_framework.tex}
\input{Sections/4_experiments.tex}
\input{Sections/5_conclusion.tex}

\bibliographystyle{ACM-Reference-Format}
\bibliography{Portrait_ST}

\input{Sections/6_sup.tex}

\end{document}

%% file: macros.tex
\usepackage{ifthen}

\definecolor{FanColor}{rgb}{0.8,0,0.8}
\definecolor{CongColor}{rgb}{0.8,0.1,0}
\definecolor{YongColor}{rgb}{0.9,0.5,0.3}
 
\newcommand{\fan}[1]{{\color{FanColor}[Fan: #1]}}
\renewcommand{\cong}[1]{{\color{CongColor}[Cong: #1]}}
\newcommand{\yong}[1]{{\color{YongColor}[Yong: #1]}}

\newcommand{\warning}[1]{{\it\color{red} #1}}
\newcommand{\toremove}[1]{{\it\color{red} (To remove) #1}}
\newcommand{\note}[1]{{\it\color{blue} #1}}
\newcommand{\nothing}[1]{}

\ifthenelse{\equal{\final}{1}}
{
\renewcommand{\cong}[1]{}
\renewcommand{\fan}[1]{}
\renewcommand{\yong}[1]{}

\renewcommand{\warning}[1]{}
\renewcommand{\toremove}[1]{}
\renewcommand{\note}[1]{}
\renewcommand{\nothing}[1]{}
}{}

%% file: Sections/authors.tex
\author{Cong Wang}
\affiliation{%
  \institution{School of Mathematics, Jilin University}
  \city{Changchun}
  \state{}
  \country{China}
}
\email{cwang16@mails.jlu.edu.cn}

\author{Fan Tang}
\affiliation{%
  \institution{SAI, Jilin University}
  \city{Changchun}
  \state{}
  \country{China}
  }
\email{tfan.108@gmail.com}

\author{Yong Zhang}
\affiliation{%
  \institution{Tencent AI Lab}
  \city{}
  \country{China}
}

\author{Weiming Dong}
\affiliation{%
 \institution{NLPR, Institute of Automation, Chinese Academy of Sciences}
 \streetaddress{}
 \city{Beijing}
 \state{}
 \country{China}}

\author{Tieru Wu}
\affiliation{%
  \institution{School of Mathematics, Jilin University}
  \streetaddress{}
  \city{Changchun}
  \state{}
  \country{China.}}

%% file: Sections/0_abstract.tex
\begin{abstract}
  Regional facial image synthesis conditioned on semantic mask has achieved great success using generative adversarial networks. 
  However, the appearance of different regions may be inconsistent with each other when conducting regional image editing.
  In this paper, we focus on the problem of harmonized regional style transfer and manipulation for facial images.
  The proposed approach supports regional style transfer and manipulation at the same time.
  A multi-scale encoder and style mapping networks are proposed in our work. 
  The encoder is responsible for extracting regional styles of real faces. 
  Style mapping networks generate styles from random samples for all facial regions.
  As the key part of our work, we propose a multi-region style attention module to adapt the multiple regional style embeddings from a reference image to a target image for generating harmonious and plausible results. 
  Furthermore, we propose a new metric ``harmony score'' and conduct experiments in a challenging setting: three widely used face datasets are involved and we test the model by transferring the regional facial appearance between datasets. Images in different datasets are usually quite different, which makes the inconsistency between target and reference regions more obvious. 
  Results show that our model can generate reliable style transfer and multi-modal manipulation results compared with SOTAs.
  Furthermore, we show two face editing applications using the proposed approach.
\end{abstract}

\ccsdesc[500]{Computing methodologies~Image representations}

\keywords{semantic synthesis, face manipulation, style transfer}

%% file: Sections/1_introduction.tex
\section{Introduction}
\input{Figures/intro_samples.tex}
Semantic image synthesis~\cite{pix2pix,bicyclegan,crn,pix2pixhd,spade,groupdnet,sean} that aims to generate realistic natural images from semantic labels is an active research topic in the past few years. 
Based on the difference in the way of involving new styles for synthesis, there are two types of mainstream methods to generate diverse images: injecting random noise~\cite{pix2pix,dsgan,bicyclegan} or transfer from referenced images~\cite{mask2face,maskgan,sean,exampleguide}.  
Researchers have made great progress in both fields.
Choi \textit{et al}.~\cite{starganv2} employ a style extraction net for facial style transfer and a mapping network adapted from StyleGAN~\cite{stylegan,stylegan2} to transform Gaussian noise into style codes. 
SPADE~\cite{spade} adopts the idea of VAE~\cite{vae} to encode the image style and enables both tasks. However, SPADE is just able to transfer facial style globally, thus limiting practical usage.
Recent works~\cite{mask2face,sean,groupdnet} propose to extract style codes of all semantic components separately, enabling regional style transfer and manipulation (R-ST\&M) for facial images.

\input{Figures/framework}

R-ST\&M provides a flexible way for facial image editing.
However, new problems arise at the same time: regional appearance editing (i.e., transfer or manipulation) will lead to the appearance of different regions inconsistent with each other.
For example, when transferring nose style from a target facial image to another with a different skin color, the new nose style of the generated result will be abrupt to the rest regions in target image (Figure~\ref{teaser}). 
Similar problems have been realized in the field of image composition~\cite{tip_harmony,dovenet,dih,zhu_harmony}. 
To the best of our knowledge, there are no prior works that focus on style consistency and harmony for R-ST\&M.
 
Generally speaking, R-ST\&M approaches should only edit the appearance of specified region(s) and the rest regions remain the same as inputs.
A harmonized R-ST\&M model should avoid affecting the appearance of remaining regions and adapt the synthetic style to the environment in target image while preserving the tone and texture information.
In this paper, we propose a framework which takes style consistency of different regions into consideration for R-ST\&M. 
We design a multi-scale encoder which incorporates feature maps from all original layers in SEAN encoder to extract style codes with richer style information, since low-level features are important for reconstruction \cite{pix2pix,bicyclegan,encoding_stylegan}.

In order to make the generated image with transferred style look plausible in synthesized image, we employ a multi-region style attention (MRSA) module where the relevance between the reference and target image is computed to synthesize a calibrated reference style. 
Apart from regional style transfer, we employ style mapping networks to map random vectors to the style spaces for region-wise multi-modal style synthesis. 
The idea of style mapping networks is inherited from StarGAN-v2 \cite{starganv2}. 
Differently, instead of training the mapping networks by adversarial loss of fake/real images, we calculate the adversarial loss on the style embedding space.
The multi-scale encoder outputs multi-region style spaces with relationship among different regions, building the mapping networks  directly from distributions can generate reliable regional styles.
To further evaluate ``harmony'' of synthetic images, we propose the metric harmony score (HS) by using a classification network to distinguish natural photographs from composite images. 

To summarize, our main contributions are as follows:
\begin{itemize}
\item We focus on the appearance harmony among regions for R-ST\&M tasks and introduce a multi-scale encoder that incorporate low- and high-level features to extract regional styles and style mapping networks to generate random styles for different semantics.
\item We introduce a multi-region style attention module which facilitates harmony and consistency in regional style transfer.
\item We conduct sufficient evaluations and show two new face editing applications to proof that the proposed framework can generate high quality facial images on various R-ST\&M tasks.  
\end{itemize}

%% file: Figures/intro_samples.tex
\begin{figure}[t]
\centering
\subfigure[Target]{
\begin{minipage}[t]{0.25\linewidth}
\centering
\includegraphics[width=0.9\linewidth]{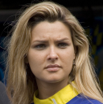}
\end{minipage}%
}%
\subfigure[Reference]{
\begin{minipage}[t]{0.25\linewidth}
\centering
\includegraphics[width=0.9\linewidth]{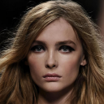}
\end{minipage}%
}%
\subfigure[Sean]{
\begin{minipage}[t]{0.25\linewidth}
\centering
\includegraphics[width=0.9\linewidth]{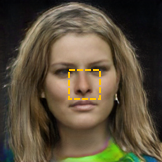}
\end{minipage}
}%
\subfigure[Ours]{
\begin{minipage}[t]{0.25\linewidth}
\centering
\includegraphics[width=0.9\linewidth]{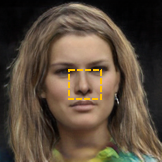}
\end{minipage}
}%
\vspace{-5mm}
\caption{An example of nose transfer (zoom in for details). R-ST\&M method can modify the nose in target image according to the appearance of the nose in reference image. However, without considering the relationship between different regions, the synthesis region in (c) is not harmonized with the other regions.}
\label{teaser}
\end{figure}


%% file: Figures/framework.tex
\begin{figure*}[t]
    \centering
    \includegraphics[width=0.9\textwidth]{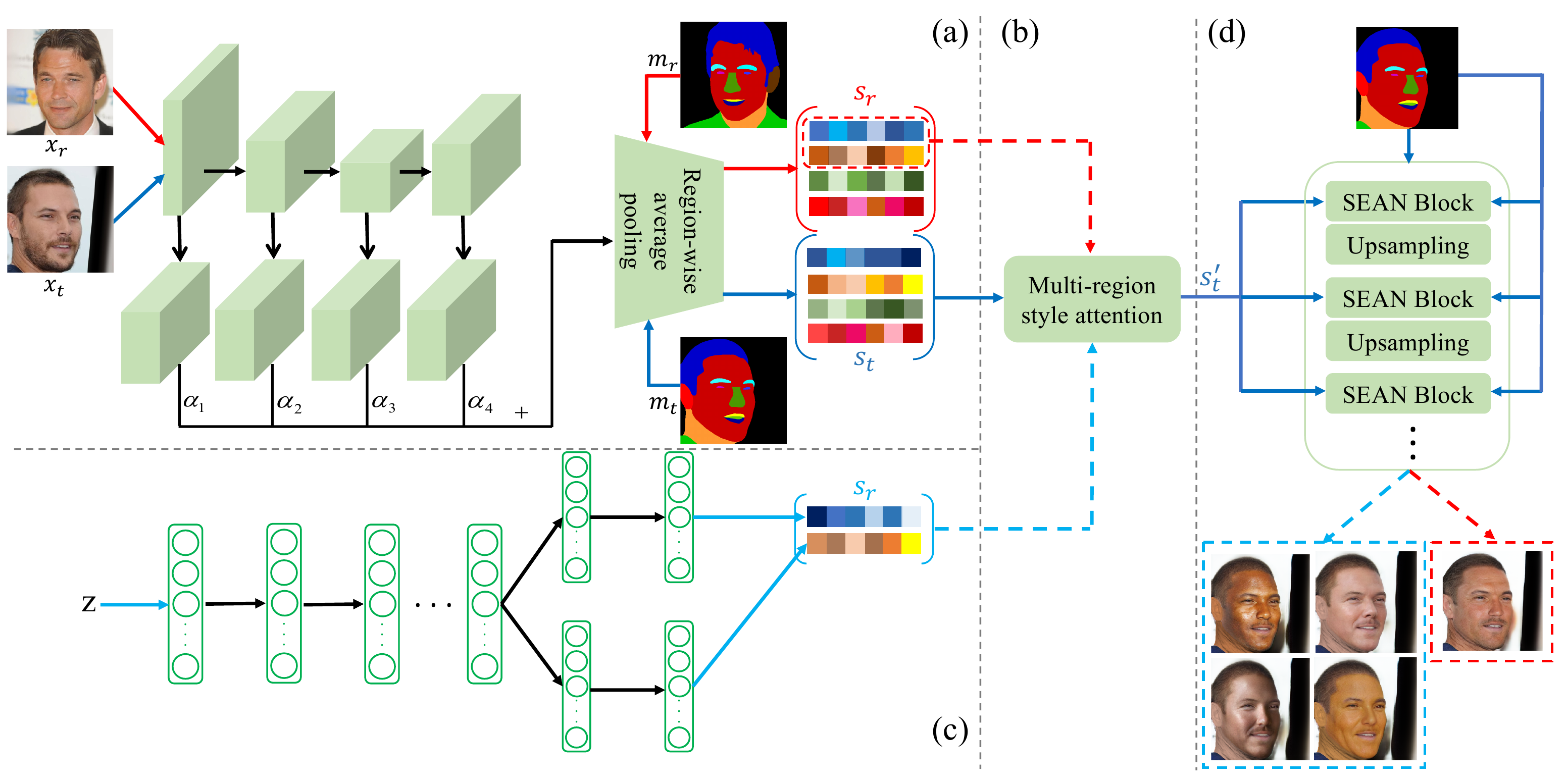}
    \caption{Whole framework of our model. (a) Multi-level feature fusion part of the encoder. (b) Multi-region style attention module. (c) An example of style mapping networks. In this example, the mapping network generates styles of skin and nose simultaneously, and multi-modal results are shown in (d). (d) SEAN generator and results of (a) and (c).}
    \label{network}
\end{figure*}

%% file: Sections/2_related.tex
\section{Related Work}

\paragraph{Facial Image Manipulation with GANs.}
Generative adversarial nets (GANs)~\cite{gan,pix2pix,wgan,stylegan,stylegan2,pggan} have achieved great success in image generation. 
A GAN consists of two competitors, \textit{i.e.,} a generator and a discriminator.
The generator is trained to synthesize images that cannot be distinguished from real ones by the discriminator.
However, the original GAN~\cite{gan} suffers from mode collapse. Then lots of works are proposed to improve the generation quality of GANs, such as \cite{laplacian-gan,wgan,gp-wgan,lsgan,sagan}. 

One of the most important applications of GANs is to generate photo-realistic human face images. 
PGGAN~\cite{pggan} is proposed to grow both the generator and discriminator progressively, allowing users to produce high-resolution and high-quality face images. 
StyleGAN~\cite{stylegan} and StyleGAN2~\cite{stylegan2} introduce a novel generator architecture borrowed from style transfer literature, enabling indistinguishable face images generation.
In the field of facial image editing, significant progress has been made using powerful GANs. 
InterFaceGAN~\cite{interfacegan} explores the disentanglement between various semantic attributes and edits several attributes using linear editing path. 
SeFa~\cite{sefa} proposes a general closed-form factorization method for latent semantic discovery. 
StyleRig~\cite{stylerig} proposes to provide a face rig-like control over a pretrained StyleGAN. 
StyleFlow~\cite{styleflow} presents to utilize normalizing flows \cite{nf} for facial attributes editing interactively with StyleGAN. 

Recent works~\cite{idinvert,encoding_stylegan} learn to encode facial images for StyleGAN inversion and facilitate various image editing tasks. 
MaskGAN~\cite{maskgan} proposes a face dataset with fine-grained mask annotations and dense mapping network for attribute transfer and style copy. 
However, MaskGAN just allows global style transfer. 
Gu et al.~\cite{mask2face} proposes an end-to-end framework to learn conditional GANs guided by semantic masks, enabling facial regional style transfer. 
SEAN~\cite{sean} proposes semantic region-adaptive normalization for GANs conditioned on segmentation masks, and the model can control the style of each semantic region individually. 
Our work improves the SEAN encoder with a multi-scale structure and a multi-region style attention module for facial image harmonization. 
Moreover, we introduce style mapping nets to generate multi-modal styles regionally with latent codes sampled from Gaussian distribution.

\paragraph{Self-Attention.}
Self-attention is first proposed in the natural language processing literature by Transformer~\cite{transformer}. 
Then computer vision researchers extend the idea to video classification~\cite{nonlocal} and image generation~\cite{sagan}. Recent works generalize self-attention to extract the correspondence between source image and reference image for semantic style transfer~\cite{cocosnet,refsketch} and makeup transfer~\cite{psgan}. 
However, the self-attention mechanism computes the correspondence spatially, making it time-consuming and inefficient. 
Differently, our style attention inspired by the above works computes the correlation among semantic regional style vectors, which ensures its computation efficiency.

\paragraph{Multi-Modal Image Synthesis.}
BicycleGAN~\cite{bicyclegan} models a distribution of possible outputs in a conditional generative modeling setting. 
To ensure that random sampling can be used during testing, the model employs KL-divergence loss to enforce the latent style distribution to be close to a standard normal distribution. \cite{munit,drit} extend the idea of multi-modal to unsupervised image-to-image translation and generate diverse images. 
SPADE~\cite{spade} uses the same idea to encode image style for semantic image synthesis. 
GroupDNet~\cite{groupdnet} extends SPADE by using KL loss for all the semantic labels, thus enabling regional multi-modal synthesis. 
Recently, StarGAN-v2~\cite{starganv2} is proposed to learn a mapping network to achieve diversity, and our style mapping model is the same as StarGAN-v2 but with different training strategy which is more suitable for our framework.

%% file: Sections/3_framework.tex
\section{Framework Architecture}
Figure~\ref{network} shows the framework of the proposed multi-region style transfer and multi-modal synthesis method. 
The inputs are a target image $x_t$ that the user wants to edit with segmentation and a reference style.
The reference style can either be generated from a reference \textit{style image} $x_r$ with segmentation for style transfer, or directly sampled from a normal Gaussian distribution for manipulation.
In this section, we start from introducing the regional feature encoding, including a multi-scale encoder for input images and regional style mapping (\textbf{RSM}) subnets for multi-modal style synthesis. 
We then move on to the multi-region style attention (\textbf{MRSA}) module followed by a semantic region-adaptive normalization based decoder.
Finally, we discuss the supervised training strategy and details. 

\input{Figures/SAM}

\subsection{Regional Feature Encoding.}
\paragraph{Multi-scale Encoder.}
The encoder in SEAN employs a ``bottleneck'' structure with plain convolutional layers to extract styles of all facial semantic regions. 
Since the purpose of the model is to generate images from the output of encoder, the low-level features from shallow layers are important for image reconstruction. 
Therefore, we compute the weighted summation of feature maps from all layers in encoder, as shown in Figure~\ref{network}(a). 
Concretely, we first re-scale the feature maps to a unified resolution and get new features $\{F_i\}_{\i=1}^K$, where $K$ is the number of shallow layers. 
Then, a set of learnable parameters $\{a_i\}_{i=1}^K$ are defined and we feed them into a softmax function for normalization as:
\begin{equation}
 \label{eq1}
  \{\alpha_i\}_{i=1}^K \leftarrow \textrm{softmax}(\{a_i\}_{i=1}^K).
\end{equation}
After that, we get the final multi-scale style feature map,
\begin{equation}
 \label{eq2}
  F = \sum_{i=1}^K{\alpha_i F_i}.
\end{equation}
The learned weights $\{a_i\}_{i=1}^K$ indicate the proportion of each part for compositing the feature map $F$. 
Given an input target image $x_t$ and a reference image $x_r$ with their segmentation masks ($m_t$ and $m_r$), we employ the region-wise average pooling layer~\cite{pix2pixhd,sean} to transform $F_t$ and $F_r$ to initial style vectors $s_t$ and $s_r$ respectively.

\paragraph{Regional Style Mapping.}
In order to synthesize multi-modal facial images with random styles, we utilize a series of regional style mapping sub-networks to learn the distributions of styles from different facial regions respectively. 
We divide all facial semantic regions into several groups according to their relevance, and one network is responsible for one group. 
For example, some regions such as skin and nose that share same color and texture appearance are strongly correlated, so we should define one network to model them simultaneously. 
As the correlations among some regions such as nose and hair are weak, we use two networks to model them separately. 
Figure~\ref{network}(c) shows an example of the mapping sub-network for modeling skin and nose.
Given a latent code $z$ sampled from the Gaussian distribution, a random reference style can be generated by
\begin{equation}
 \label{eq2}
  s_r = \mathcal{M}(z).
\end{equation}
In our method, related regions such as skin and nose or two eyes share a same mapping network.
More details of the training of RSM are in Sec.~\ref{sec:model_training}.
After that, we can feed $s_r$ into the MRSA module and $\mathcal{G}$.


\subsection{Multi-region Style Attention}
If the global appearances (i.e., lighting conditions) in $x_t$ and $x_r$ are quite different, regional style transfer results probably become inharmony. 
However, users prefer to get a harmonious image directly without subsequent image harmonization process. 
To this end, we propose a multi-region style attention (MRSA) module to learn transferred styles.
Figure~\ref{sa} illustrates the workflow of MRSA. 
Different from the attention modules in \cite{cocosnet} and \cite{refsketch} that extract the spatial correspondence in pixel space, our MRSA module computes the relevance of regional semantic styles. 
In order to correct the styles of different regions, we first concatenate the corresponding components in $s_r$ with the rest components in $s_t$ to form a new $s^{\prime}_r$.
Then we map the style vectors using $Q = \mathcal{W}_q(s^{\prime}_r)$, $K = \mathcal{W}_k(s_t)$ and $V = \mathcal{W}_v(s_t)$, where $\mathcal{W}_q$, $\mathcal{W}_k$ and $\mathcal{W}_v$ are linear mappings. 
After that, an attention matrix can be computed by $Q*K^\top$ followed by a softmax function within each row, \textit{i.e.,}
\begin{equation}
 \label{eq3}
  M = {\rm softmax} (Q*K^\top),
\end{equation}
where * denotes matrix multiplication. 
After computing the attention matrix $M$, we can get the style correction $s_c=M*V$. 
Finally, the target style can be computed by
\begin{equation}
s^{\prime}_t = s^{\prime}_r + \alpha s_c.
\end{equation}

\subsection{Decoder.}
Given the style vectors generated by MSRA, the SEAN generator~\cite{sean} is used as decoder by feeding them into a semantic region-adaptive normalization (SEAN) module. 
In the SEAN normalization, target mask along with style map generated by broadcasting style vectors to the corresponding regions are used to modulate the activation from previous layer. 
The decoder employs several SEAN blocks with upsampling layers and synthesizes images progressively.

\subsection{Model Training}
\label{sec:model_training}
The encoder-decoder part in our model is similar with SPADE and SEAN.  
We use three loss functions described in SPADE and SEAN to train this part: adversarial loss, feature matching loss and perceptual loss.
During training, if we use the $s_r$ extracted from a reference different from the source image, this results in an unsupervised training as there is no ground truth for the new image. 
To tackle this problem, we set $x_r$ equal to $x_s$ for training. 
We test the training strategy by mixing supervised with unsupervised training, but it fails to generate realistic images. 
The reason we suppose is that the unsupervised result would disturb supervised training pace.

As for style mapping networks $\{\mathcal{M}_j\}_{j=1}^M$, we turn to the adversarial loss imposed on $s_s$ and $s_r$ generated by style mapping. 
$M$ is the number of mapping networks.
In order to  train $\{\mathcal{M}_j\}_{j=1}^M$, a set of discriminators $\{\mathcal{D}_j\}_{j=1}^M$ are employed and the adversarial objectives are as follows:
\begin{equation}
 \label{eq4}
  \mathcal{L}_j = \min_{\mathcal{M}_j} \max_{\mathcal{D}_j} \mathbb{E}[\log{\mathcal{D}_j(s_s)}] + \mathbb{E}[\log{(1-\mathcal{D}_j(\mathcal{M}_j(z)))}].
\end{equation}
A similar style mapping network has been proposed in StarGAN-v2 \cite{starganv2} which focuses on unsupervised image-to-image translation. 
However, StarGAN-v2 trains it with the adversarial loss defined on image synthesis. 
The training strategy in StarGAN-v2 cannot effectively train our style mapping networks. 
The reason is that our encoder-decoder part is trained in a supervised way, and the encoder will learn expressive style information. 
It is more effective to learn the distributions of encoded styles directly.


%% file: Figures/SAM.tex
\begin{figure}[t]
    \centering
    \includegraphics[width=\linewidth]{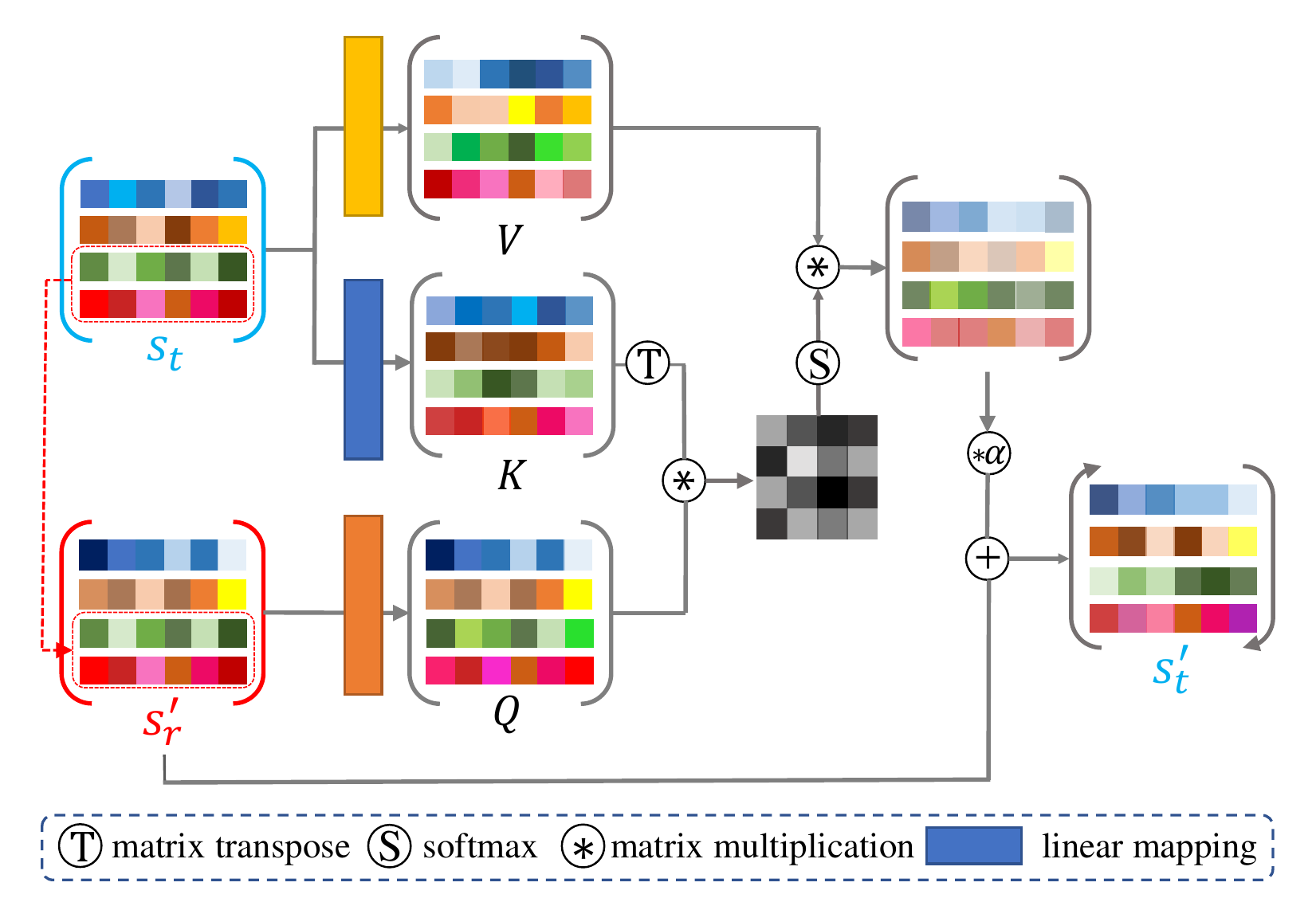} 
    \caption{Multi-region style attention module. $s_t$ denotes target style vectors of all regions. $s^{\prime}_r$ is the concatenation of styles from target regions in the reference and styles from the rest regions in the target. Linear projection metric $\mathcal{W}_v$, $\mathcal{W}_q$ and $\mathcal{W}_k$ are adopted to produce $V$, $Q$ and $K$, respectively. Then, $V$ and $Q$ are used to yield an attention matrix $M$. Finally, the multi-region style correction is calculated by $M*V$ which is applied to $s^{\prime}_r$.}
    \label{sa}
\end{figure}

%% file: Sections/4_experiments.tex
\section{Experiments}
\subsection{Experimental Settings.}
\paragraph{Datasets.} We use three face datasets to evaluate our framework:
\begin{itemize}
\item CelebAMASK-HQ~\cite{maskgan} consists of $30,000$ face images with segmentation masks.
Each image is annotated with a semantic mask of $19$ semantic categories in total. 
We use the first $28,000$ images for training and the remains for evaluating.
\item FFHQ~\cite{stylegan} contains $70,000$ high-quality images. 
We utilize a deeplab-v3 model \cite{deeplabv3} trained on CelebAMASK-HQ to parse the facial semantics. 
We employ the first $2,000$ images for evaluation. 
\item LaPa~\cite{lapa} is a new dataset for face parsing which consists of more than $22,000$ images with large variations in pose, facial expression and illumination. 11-category semantic label maps are provided. 
We discard low-resolution images in the dataset. 
The final training set contains $19,770$ faces and testing set contains $1,930$ faces.
\end{itemize}
\input{Tables/reconstruction}
\paragraph{Metrics.} We employ several commonly used metrics to evaluate our framework and the competing state-of-the-art methods. 
Specifically, FID~\cite{ttur} computes the distance between the distributions of synthesized images and the distribution of real images, which is used to evaluate the quality of synthesized results. 
We also adopt PSNR, SSIM and LPIPS~\cite{lpips} to assess the similarity between the synthesized and the ground-truth image in face reconstruction task. 
In order to evaluate the performance of our model for regional multi-modal synthesis with random styles, we utilize mean Class-Specific Diversity (mCSD) and mean Other-Classes Diversity (mOCD)~\cite{groupdnet}. 
For a fixed semantic region, mCSD is used to assess the generation diversity of the region while mOCD is used to assess the diversity of the rest regions.
Apparently, high mCSD and low mOCD indicate good performance for the fixed region. 

In addition to the above existing metrics, we propose a new metric, named \textbf{harmony score} (HS), to measure the harmony degree between the transferred region and the rest for regional style transfer. 
Concretely, we train a convolutional neural network to distinguish real images from synthetic ones and use the output probability as the harmony score. 
The real images are set as positive samples and the unrealistic composite images are set as negative samples. 
We use HAdobe5k~\cite{dovenet} to train the classification network and concatenate one image and the corresponding foreground mask as an input. 

\paragraph{Competing methods.}
We compare our method with four leading semantic image synthesis models: pix2pixHD~\cite{pix2pixhd}, SPADE~\cite{spade}, GroupDNet~\cite{groupdnet} and SEAN~\cite{sean}. 
Specifically, pix2pixHD applies an image feature encoder network and instance-wise pooling to get image features within each object. Then, the features and the corresponding mask are feed into a coarse-to-fine generator to reconstruct the image. 
Therefore, pix2pixHD is suitable for regional style transfer. 
SPADE proposes the encoder and generator to form a VAE~\cite{vae} and a new normalization for the generator, enabling global style transfer and multi-modal synthesis conditioned on semantic mask. 
GroupDNet extends the idea of SPADE by encoding different semantic regions separately and leveraging group decreasing generator. 
GroupDNet can be used for regional style transfer and multi-modal synthesis. 
SEAN employs similar structures of pix2pixHD encoder and SPADE generator, and it improves the generation quality significantly with the SEAN normalization. 
\input{Figures/reconstruction}
\subsection{Implementation Details}
We use the TTUR~\cite{ttur} strategy and set the learning rate to $0.0001$ and $0.0004$ for the generator and discriminator, respectively.
Following SPADE and SEAN, we apply Spectral Norm \cite{sn} to the encoder.
Moreover, we use the ADAM solver~\cite{adam} with $\beta_1=0.5$ and $\beta_2=0.999$ to optimize the model. For style mapping, we set the learning rate to 0.0002 for both mapping networks and discriminators. For both training and evaluation, the input images are resized to a fixed resolution of 256 $\times$ 256. 


\subsection{Global Reconstruction}
We first evaluate the effectiveness of the proposed multi-region style control and manipulation network in image reconstruction task, namely transferring the own style to itself. 
Only one image is employed as input.
Visual comparisons are shown in Figure~\ref{recon}.
Overall, pix2pixHD and groupDNet cannot maintain the skin color well. 
Compared with SEAN, our method can reconstruct more facial details of the input, \textit{e.g.} the wrinkles on the left side of the woman's face and the left eye under sunglasses of the man.
In terms of quantitative evaluation, as shown in Table~\ref{tab1}, our model outperforms other SOTA methods on all datasets. 
It is worth mentioning that although MRSA is designed for style transfer and manipulation, it exhibits the best reconstruction quality (\textit{i.e.,} the lowest FID) on all datasets.

\subsection{Regional Style Transfer}
\label{harmony}
We further evaluate the effectiveness of the proposed approach in regional style transfer task.
One target image and one reference image are employed as inputs.
We split all testing datasets into two parts: one half as target images and the other half as reference images. 
SPADE is not selected for comparison since it does not support region style transfer.
Figure~\ref{regionST} (a) shows the quantitative results.
We use FID as the metric in two transfer tasks: skin (with nose) transfer and hair transfer. 
The quantitative results are shown in Table~\ref{tab2}. 
In terms of FID, our model with style attention achieves the lowest values, indicating that it synthesizes human faces with the highest quality.

\input{Figures/styletransfer}

\input{Tables/styletransfer}

\begin{table*}[t]
  \caption{Results (PSNR$\downarrow$/HS$\uparrow$) for regional style transfer across datasets.}
  \label{tab3}
  \centering
  \scalebox{1}{
  \begin{tabular}{l|cc|cc|cc}
  \hline
    & \multicolumn{2}{c|}{CelebAMASK-HQ$\rightarrow$LaPa} & \multicolumn{2}{c}{FFHQ$\rightarrow$LaPa} & \multicolumn{2}{|c}{LaPa$\rightarrow$LaPa} \\
    & skin & hair & skin & hair & skin & hair \\
  \hline
  pix2pixHD \cite{pix2pixhd}   & 73.62/0.7923 & 73.80/0.8075 & 77.36/0.8538 & 73.98/0.8137 & 85.49/0.8314 & 84.72/0.8035 \\
  GroupDNet \cite{groupdnet}   & 96.20/0.8965 & 93.36/0.8752 & 94.78/0.9131 & 93.34/0.8753 & 104.52/0.9093 & 104.02/0.8736 \\
  SEAN  \cite{sean}            & 48.25/0.7420 & 48.43/0.6940 & 48.17/0.7105 & 48.62/0.7164 & 61.19/0.7396 & 60.68/0.6996 \\
  \hline \hline
  Ours w/o softmax    & 52.98/0.8071 & 54.90/0.7353 & 53.10/0.8100 & 54.68/0.7495 & 66.67/0.8003 & 65.25/0.7348 \\
  Ours w/o SA         & 48.06/0.7749 & 47.72/0.7130 & 47.84/0.7598 & 48.39/0.7310 & 61.43/0.7872 & 59.99/0.6964 \\

  Ours                & 47.46/0.8490 & 47.05/0.7742 & 46.71/0.8341 & 47.36/0.7854 & 60.14/0.8537 & 58.85/0.7566 \\
  \hline
  \end{tabular}}
\end{table*}

\paragraph{Transfer Cross dataset.}
Most images in CelebAMask-HQ~\cite{maskgan} and FFHQ~\cite{stylegan} are captured in good lighting conditions. 
Regional style transfer among these images can barely lead to inharmony. 
However, LaPa~\cite{lapa} consists of facial images with abundant variations in lighting conditions. 
Therefore, we transfer skin (with nose) and hair of facial images in the test sets of CelebAMask-HQ, FFHQ and LaPa to the test set of LaPa separately, and calculate FID and HS of synthesized faces.
Quantitative and qualitative results are presented in Table \ref{tab3} and Figure~\ref{regionST} (b)\textasciitilde(d), respectively. 
According to the quantitative results, our method and SEAN~\cite{sean} perform much better than pix2pixHD~\cite{pix2pixhd} and GroupDNet~\cite{groupdnet} in terms of FID. 
Our model and SEAN generate regional transfer results with higher image quality. 
We can draw the same conclusion from Figure~\ref{regionST}. 
The proposed metric HS reflects the harmony degree between the transferred region and the rest regions. 
Our method exhibits obviously higher harmony score than SEAN while pix2pixHD and GroupDNet reach higher harmony scores than our model. 
As shown in Figure~\ref{regionST}, although the results of pix2pixHD and GroupDNet are harmonious, the two methods fail to reconstruct the transferred styles and the rest regions expected to keep their appearance change severely. 
In summary, our model is the best trade-off considering image quality and harmony degree.

\paragraph{User Study.}
We conduct user studies to further compare the visual performance of ours and the SOAT methods aforementioned.
Firstly, we show the participants each target-reference pair and tell them which region in the target image we want to edit.
Then we show them four results, one is by our method and the others are from pix2pixHD, GroupDNet and SEAN.
Each subject are assigned with $30$ group results.
We receive $59$ responses, among which $47$ responses are valid.
A total of $1,410$ votes are obtained. 
Our model has $627$ ($44.45\%$) votes, SEAN has $410$ ($29.07\%$) votes, GroupDNet has $265$ ($18.78\%$) votes, and Pix2Pix has $108$ ($7.66\%$) votes.

\input{Tables/multimodal}

\begin{figure*}[t]
    \centering
    \includegraphics[width=\textwidth]{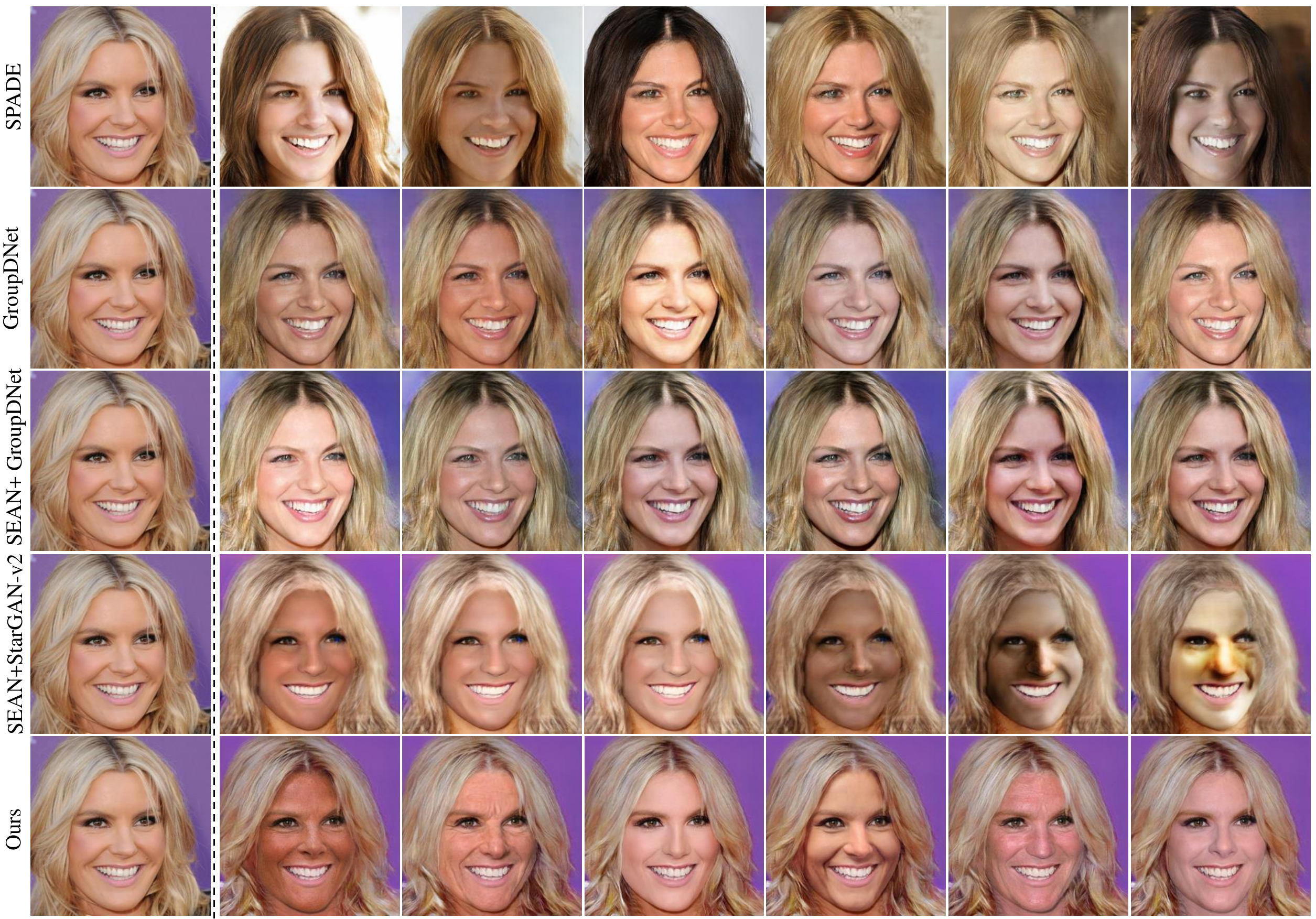}
    \caption{Skin multi-modal synthesis.}
    \label{diversity}
\end{figure*}

\subsection{Regional Multi-modal Manipulation}
\label{multi-modal}
We finally evaluate the effectiveness of the proposed approach in regional multi-modal manipulation task.
One target image and one vector sampled from a normal Gaussian distribution are employed as inputs.
SPADE~\cite{spade} and GroupDNet~\cite{groupdnet} are selected for multi-modal synthesis.
SPADE is proposed for global multi-modal synthesis while GroupDNet is for regional multi-modal synthesis.
Figure~\ref{diversity} shows the manipulation of skin, GroupDNet affects hair more significantly than ours when doing skin multi-modal synthesis. 
We further conduct qualitative experiments on manipulation of skin and hair regions. 
Table~\ref{tab4} reports the FID, mCSD and mOCD calculated over the three different datasets. 
In terms of image quality, our method outperforms the two methods by a large margin in all the datasets. 
For skin multi-modal synthesis, our method presents higher mCSD and lower mOCD. 
That is to say, our method is better at maintaining the appearance of the rest regions while achieving high color and texture diversity of skin synthesis. 
For hair multi-modal synthesis, GroupDNet generates facial images with low diversity since mCSD and mOCD of GroupDNet are close to zero. 
For all the multi-modal synthesis experiments, we manipulate each image using 10 random styles.

\begin{table}
\caption{Results (FID$\downarrow$) for ablation study of RSM.}
\label{tab5}
\centering
\begin{tabular}{cc|c|c|c} 
\hline
                                                                         &      & \begin{tabular}[c]{@{}c@{}}SEAN+\\GroupDNet\end{tabular} & \begin{tabular}[c]{@{}c@{}}SEAN+\\StarGAN-v2\end{tabular} & Ours   \\ 
\hline
\multirow{2}{*}{\begin{tabular}[c]{@{}c@{}}CelebA\\MASK-HQ\end{tabular}} & Skin & 25.32                                                    & 27.19                                                     & \textbf{12.67}  \\
                                                                         & Hair & 28.82                                                    & 20.05                                                     & \textbf{12.55}  \\ 
\hline
\multirow{2}{*}{FFHQ}                                                    & Skin & 55.02                                                    & 40.88                                                     & \textbf{31.43}  \\
                                                                         & Hair & 58.07                                                    & 33.57                                                     & \textbf{28.45}  \\ 
\hline
\multirow{2}{*}{LaPa}                                                    & Skin & 84.72                                                    & 105.30                                                    & \textbf{40.47}  \\
                                                                         & Hair & 87.29                                                    & 90.75                                                     & \textbf{41.27}  \\
\hline
\end{tabular}
\end{table}

\subsection{Ablation Study}
\paragraph{Ablation of MRSA.} To validate the effects of the softmax function and MRSA module in the encoder, we conduct ablation experiments by not using them in the framework. 
``Ours w/o softmax'' means we do not use softmax normalization and MRSA and ``Ours w/o SA'' means we do not use MRSA in our framework. 
Results of cross-dataset regional style transfer in Table~\ref{tab3} indicate that softmax normalization and MRSA improves image quality and harmony degree, respectively.

\paragraph{Ablation of RSM.} To validate the effects of the RSM module, we conduct ablation experiments by combining the encoders and training strategies used in GroupDNet~\cite{groupdnet} and StarGAN-v2~\cite{starganv2} with the SEAN generator, respectively. 
We can get two variations for comparisions, \textit{i.e.}, ``SEAN+GroupDNet'' and ``SEAN+StarGAN-v2''. 
Table~\ref{tab5} illustrates FID of skin and hair multi-modal synthesis on all the datasets. 
Our method performs much better that the two variations in terms of image quality. 
We use the same mapping network as StarGAN-v2 but different training strategy. 
If we use the training strategy in StarGAN-v2, the generator would be trained in an unsupervised way. 
However, our encoder-decoder part is trained in a supervised way and the mapping network and the encoder share the same generator. 
Thus, different objectives would misguide the generator. 
Visual comparisons can be found in Figure~\ref{diversity}.

\section{Applications}
Our framework can enable various applications in facial image synthesis. 
Sections~\ref{harmony} and~\ref{multi-modal} demonstrate the effectiveness of regional style transfer cross facial images and multi-modal synthesis with random styles, respectively. We now introduce other two applications of interactive face editing. 

\begin{figure}[th]
\centering
\includegraphics[width=\linewidth]{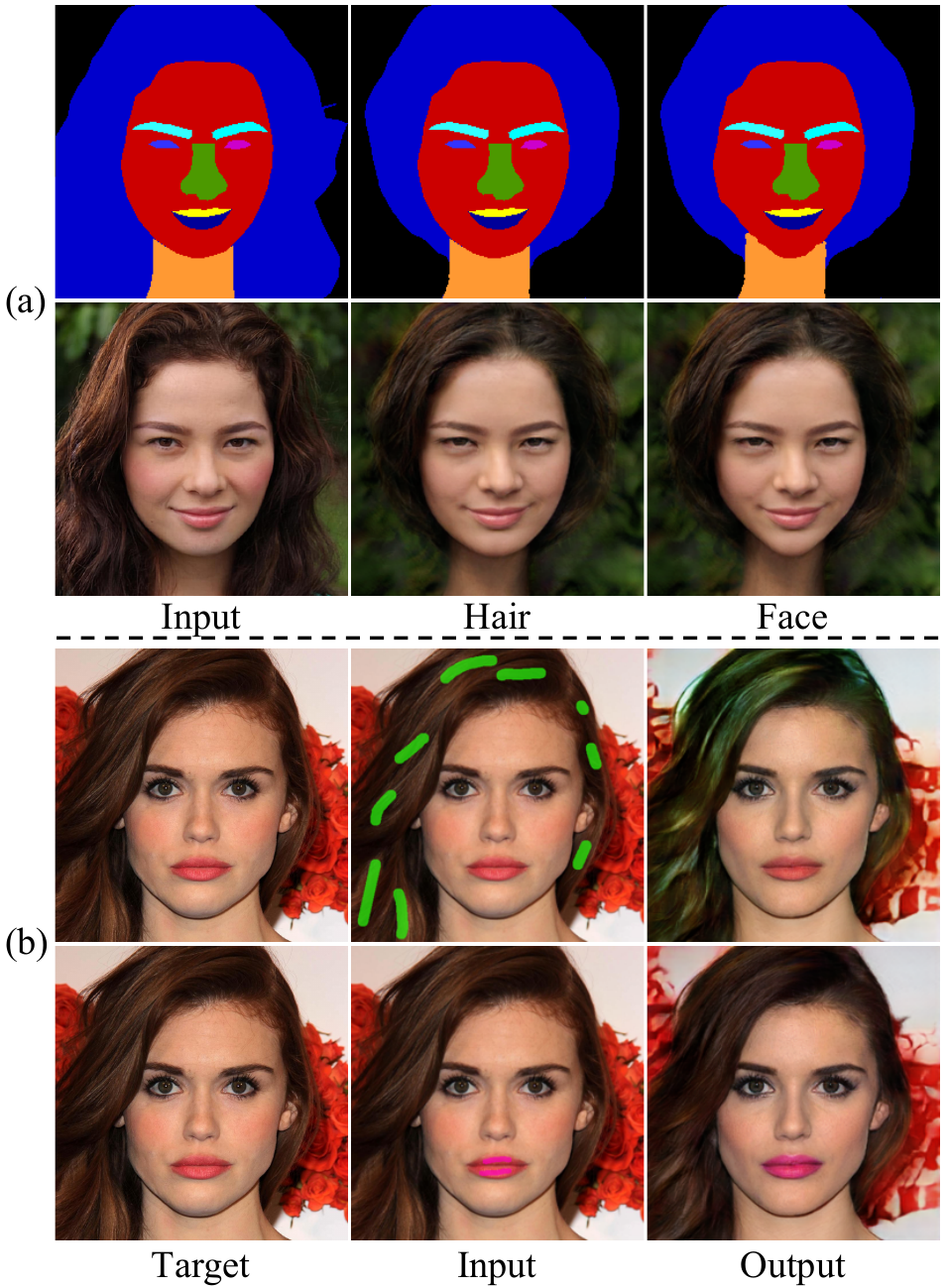}
\caption{Applications (zoom in for details). (a) Shape editing. (b) Color editing (hair/lips).}
\label{applications}
\end{figure}

\paragraph{Shape editing.} 
Our framework allows users to edit the shape of facial components directly on segmentation mask to manipulate face interactively. Figure~\ref{applications} (a) shows an example of hair and face shape editing.

\paragraph{Color editing.} 
By drawing simple color strokes on facial components, our method enables color editing on facial semantic regions. The two rows in Figure~\ref{applications} (b) demonstrate hair and lips color editing, respectively.

%% file: Tables/reconstruction.tex
\begin{table*}[ht]
  \caption{The results of facial image reconstruction. For PSNR and SSIM, the higher the better. For LPIPS and FID, the lower the better.}
  \label{tab1}
  \centering
  \scalebox{0.95}{
  \begin{tabular}{l|cccc|cccc|cccc}
  \hline
    & \multicolumn{4}{c|}{CelebAMASK-HQ} & \multicolumn{4}{c}{FFHQ} & \multicolumn{4}{|c}{LaPa} \\
    & PSNR$\uparrow$ & SSIM$\uparrow$ & LPIPS$\downarrow$ & FID$\downarrow$ & PSNR$\uparrow$ & SSIM$\uparrow$ & LPIPS$\downarrow$ & FID$\downarrow$ & PSNR$\uparrow$ & SSIM$\uparrow$ & LPIPS$\downarrow$ & FID$\downarrow$ \\
  \hline
  pix2pixHD \cite{pix2pixhd}  & 17.32 & 0.5387 & 0.2117 & 21.68 & 16.08 & 0.5200 & 0.2506 & 45.55 & 13.16 & 0.4387 & 0.3817 & 68.87 \\
  SPADE \cite{spade}          & 16.87 & 0.5142 & 0.2462 & 25.46 & 15.82 & 0.4894 & 0.2923 & 53.10 & \bf{14.82} & 0.4607 & 0.3927 & 89.96 \\
  GroupDNet \cite{groupdnet}  & 16.40 & 0.5184 & 0.2526 & 38.87 & 15.27 & 0.4913 & 0.2981 & 71.83 & 14.22 & 0.4454 & 0.3928 & 93.35 \\
  SEAN  \cite{sean}           & 18.55 & 0.5741 & 0.1749 & 17.12 & 17.23 & 0.5368 & 0.2099 & 34.29 & 14.72 & 0.4841 & 0.3281 & 47.94 \\
  Ours        & \bf{18.60} & \bf{0.5787} & \bf{0.1702} & \bf{15.26} & \bf{17.41} & \bf{0.5510} & \bf{0.2020} & \bf{32.84} & 14.75 & \bf{0.4891} & 0.3295 & \bf{46.62} \\
  \hline
  \end{tabular}}
\end{table*}

%% file: Figures/reconstruction.tex
\begin{figure}[t]
\centering
\subfigure[Input]{
\begin{minipage}[t]{1.2cm}
\centering
\includegraphics[width=1\linewidth]{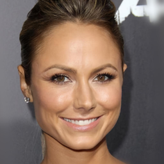}
\includegraphics[width=1\linewidth]{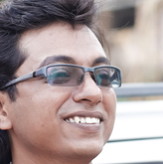}
\end{minipage}%
}
\subfigure[pix2pixHD]{
\begin{minipage}[t]{1.2cm}
\centering
\includegraphics[width=1\linewidth]{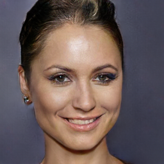}
\includegraphics[width=1\linewidth]{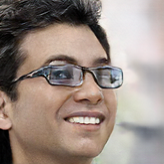}
\end{minipage}%
}
\subfigure[GroupDNet]{
\begin{minipage}[t]{1.2cm}
\centering
\includegraphics[width=1\linewidth]{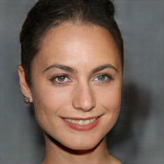}
\includegraphics[width=1\linewidth]{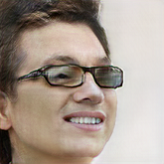}
\end{minipage}
}
\subfigure[SPADE]{
\begin{minipage}[t]{1.2cm}
\centering
\includegraphics[width=1\linewidth]{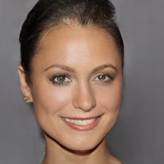}
\includegraphics[width=1\linewidth]{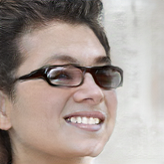}
\end{minipage}
}
\subfigure[SEAN]{
\begin{minipage}[t]{1.2cm}
\centering
\includegraphics[width=1\linewidth]{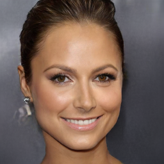}
\includegraphics[width=1\linewidth]{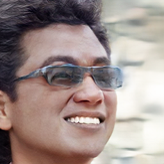}
\end{minipage}
}
\subfigure[Ours]{
\begin{minipage}[t]{1.2cm}
\centering
\includegraphics[width=1\linewidth]{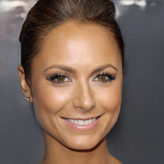}
\includegraphics[width=1\linewidth]{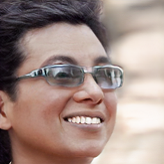}
\end{minipage}
}
\caption{Results of image reconstruction.}
\label{recon}
\end{figure}

%% file: Figures/styletransfer.tex
\begin{figure*}[t]
    \centering
    \includegraphics[width=\textwidth]{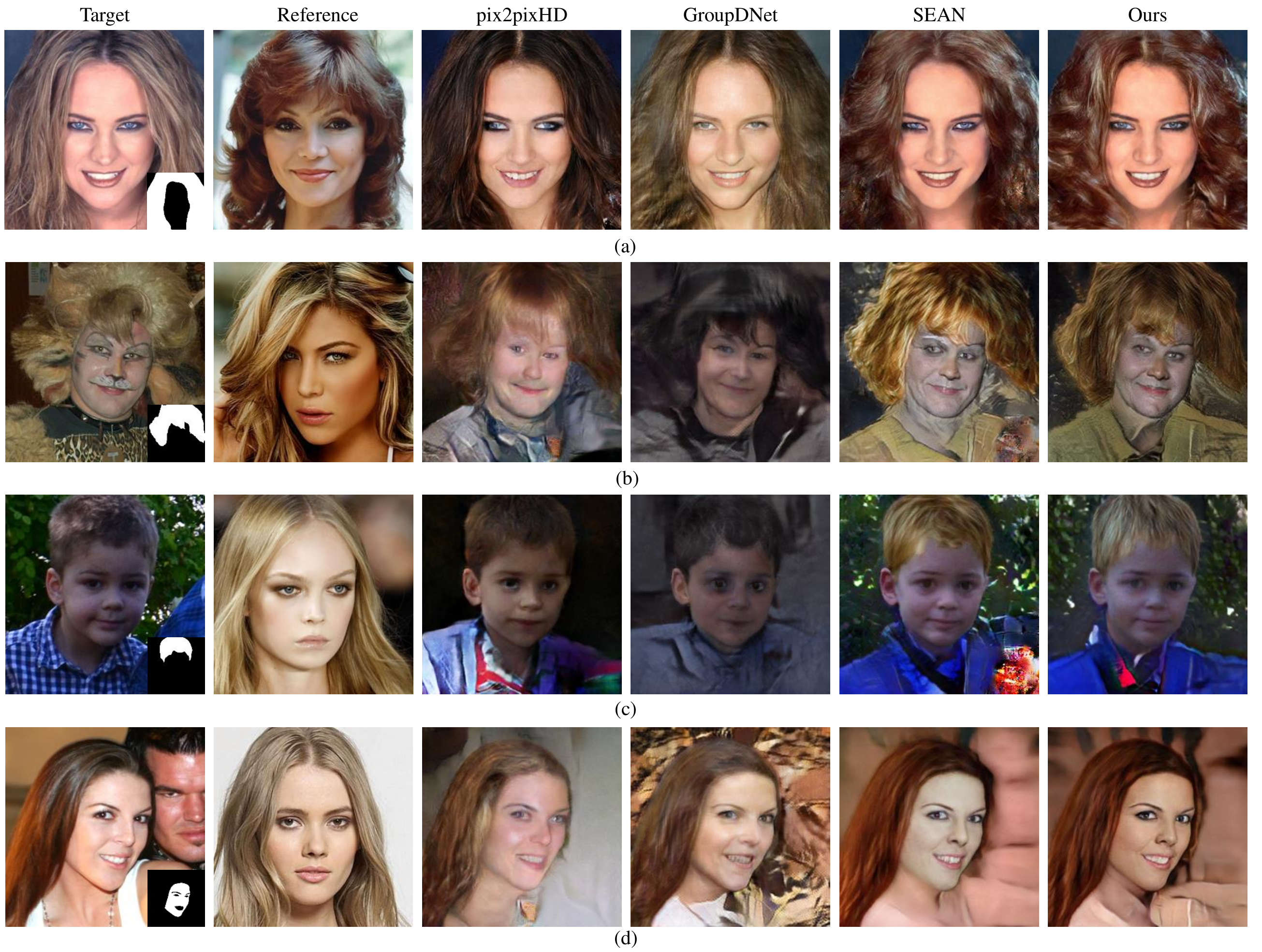}
    \vspace{-5mm}
    \caption{Results of region style transfer (segmentation mask shown as small inset). Target and reference images in (a) are from the same dataset while those in (b)\textasciitilde(d) are from different datasets.}
    \label{regionST}
\end{figure*}

%% file: Tables/styletransfer.tex
\begin{table}[t]
  \caption{The FID$\downarrow$ results of skin and hair transfer.}
  \label{tab2}
  \centering
  \scalebox{0.9}{
  \begin{tabular}{l|cc|cc|cc}
  \hline
    & \multicolumn{2}{c|}{CelebAMASK-HQ} & \multicolumn{2}{c}{FFHQ} & \multicolumn{2}{|c}{LaPa} \\
    & skin & hair & skin & hair & skin & hair \\
  \hline
  pix2pixHD         & 26.39 & 26.58 & 57.13 & 56.44 & 85.49  & 84.72 \\
  GroupDNet         & 44.71 & 44.36 & 77.89 & 78.20 & 104.52 & 104.02 \\
  SEAN              & 23.85 & 25.29 & 44.84 & 43.64 & 61.19  & 60.68 \\
  Ours              & \bf{23.00} & \bf{23.61} & \bf{42.82} & \bf{41.85} & \bf{60.14} & \bf{58.85}\\
  \hline
  \end{tabular}}
\end{table}

%% file: Tables/multimodal.tex

\begin{table}
\centering
\caption{Regional multi-modal synthesis.}
\begin{tabular}{cc|c|c|c|c} 
\hline
\multicolumn{1}{c}{}                                                     & \multicolumn{1}{c}{}  &      & SPADE  & GroupDNet & Ours    \\ 
\hline
\multirow{6}{*}{\begin{tabular}[c]{@{}c@{}}CelebA\\MASK-HQ\end{tabular}} & \multirow{3}{*}{Skin} & FID$\downarrow$  & 21.09  & 39.72     & \textbf{12.67}   \\
                                                                         &                       &mCSD$\uparrow$& 0.0354 & 0.0321    & \textbf{0.0395}  \\
                                                                         &                       & mOCD$\downarrow$& 0.2126 & 0.1280    & \textbf{0.0752}  \\ 
\cline{3-6}
                                                                         & \multirow{3}{*}{Hair} & FID$\downarrow$  & 21.12  & 50.43     & \textbf{12.55}   \\
                                                                         &                       &mCSD$\uparrow$& 0.1848 & 0.0001    & \textbf{0.2106}  \\
                                                                         &                       &mOCD$\downarrow$& 0.1230  & \textbf{0.0000}         & 0.0510   \\ 
\hline
\multirow{6}{*}{FFHQ}                                                    & \multirow{3}{*}{Skin} & FID$\downarrow$  & 51.38  & 72.34     & \textbf{31.43}   \\
                                                                         &                       &mCSD$\uparrow$& 0.0392 & 0.0360    & \textbf{0.0413}  \\
                                                                         &                       & mOCD$\downarrow$& 0.2020 & 0.0820    & \textbf{0.0279}  \\ 
\cline{3-6}
                                                                         & \multirow{3}{*}{Hair} & FID$\downarrow$  & 51.36  & 81.71     & \textbf{28.45}   \\
                                                                         &                       &mCSD$\uparrow$& 0.0723 & 0.0000         & \textbf{0.0864}  \\
                                                                         &                       & mOCD$\downarrow$& 0.1920  & \textbf{0.0000}         & 0.0140   \\ 
\hline
\multirow{6}{*}{LaPa}                                                    & \multirow{3}{*}{Skin} & FID$\downarrow$  & 74.61  & 96.75     & \textbf{40.47}   \\
                                                                         &                       &mCSD$\uparrow$& 0.0455 & 0.0446    & \textbf{0.0600}  \\
                                                                         &                       & mOCD$\downarrow$& 0.3005 & 0.1657    & \textbf{0.1071}  \\ 
\cline{3-6}
                                                                         & \multirow{3}{*}{Hair} & FID$\downarrow$  & 74.68  & 150.46    & \textbf{41.27}   \\
                                                                         &                       &mCSD$\uparrow$& 0.0512 & 0.0047    & \textbf{0.0939}  \\
                                                                         &                       & mOCD$\downarrow$& 0.3080  & \textbf{0.0000}     & 0.0980  \\
\hline
\end{tabular}
\end{table}

%% file: Sections/5_conclusion.tex
\section{Conclusion}
In this paper, we focus on the harmonized region style editing for facial images.
The proposed framework follows the encoding-fusion-decoding fashion.
For the encoder, we employ a multi-scale structure in order to extract regional styles more effectively. 
Then a multi-region style attention (MRSA) module is proposed for harmonious regional style transfer, especially when the target and reference face images are with different lighting conditions. 
For the sake of regional multi-modal synthesis, we introduce the regional style mapping (RSM) net to map random noise to styles. 

Although our model can generate high quality regional multi-modal results with random styles, the styles of specific region are still in weak control condition. 
The regional style transfer is the only way to provide strong control information.
If we want to randomly synthesize regions with specified appearance, our model, SPADE and GroupDNet will be helpless. 
This problem remains to be resolved and it will be our future work.

%% file: Sections/6_sup.tex
\appendix
\section{Supplementary Material}

\begin{figure*}[t]
\centering
\includegraphics[width=\linewidth]{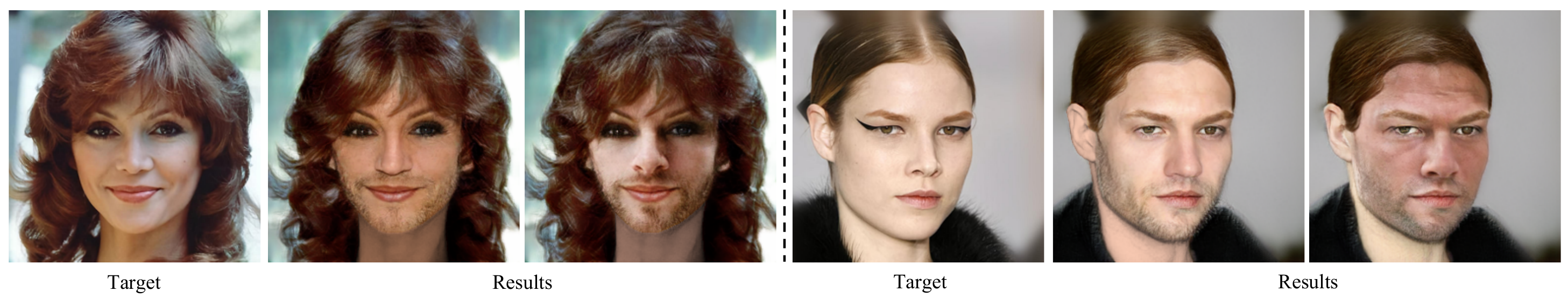}
\caption{Bad cases of our method for skin multi-modal synthesis. }
\label{bad_case}
\end{figure*}

\begin{figure*}[t]
\centering
\includegraphics[width=\linewidth]{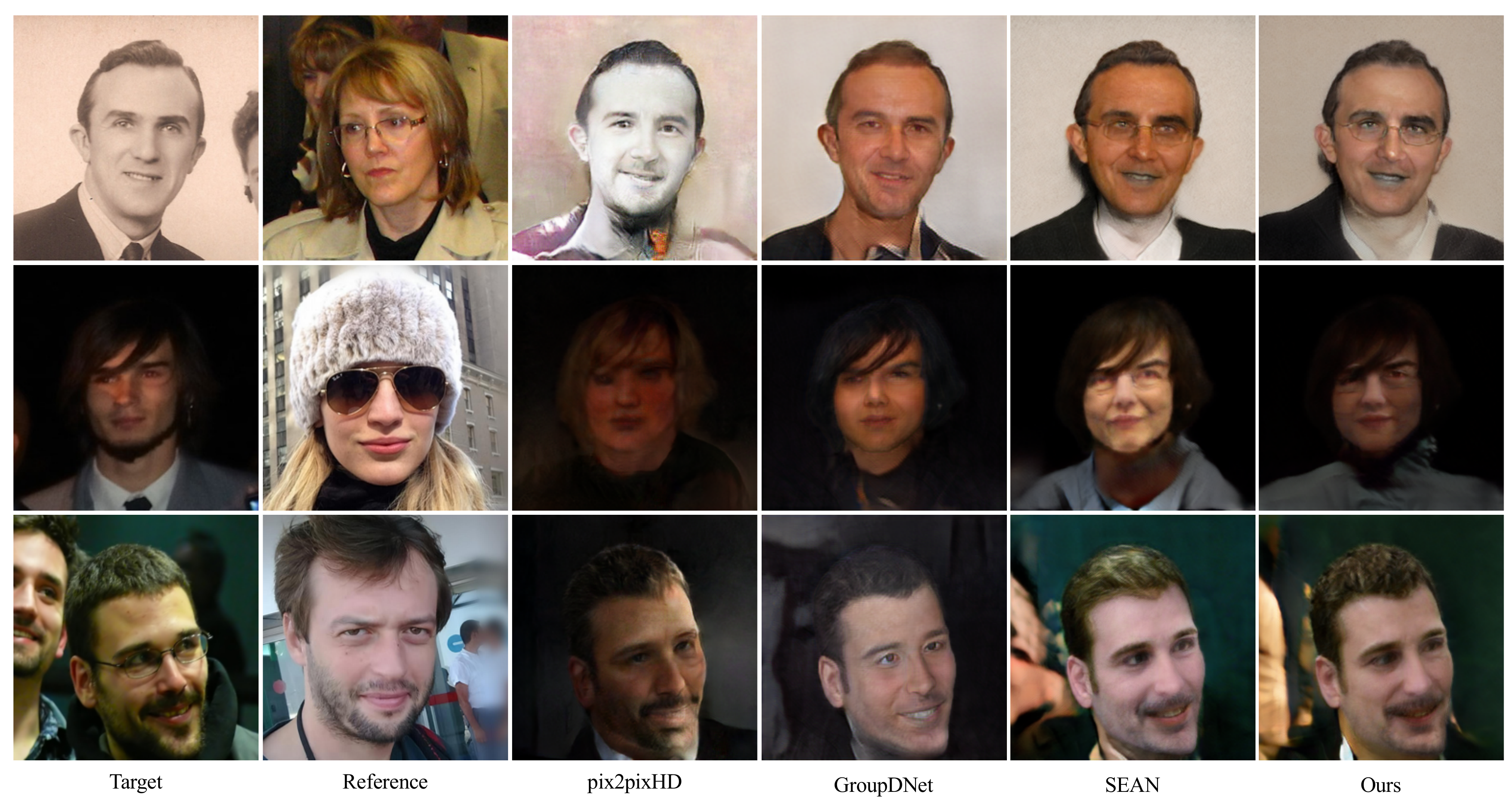}
\caption{Skin style transfer of our method and competing methods. }
\label{supp_harmony-skin}
\end{figure*}

\paragraph{Bad cases.}
Figure~\ref{bad_case} shows some cases where the proposed model failed to generate reasonable results: women have beards. This is because the proposed approach did not take the semantic of different region into consideration. Our method only consider the appearance relation among different regions. For regional multi-modal synthesis, our model cannot control the semantics of random style codes. 

\begin{figure*}[t]
\centering
\includegraphics[width=\linewidth]{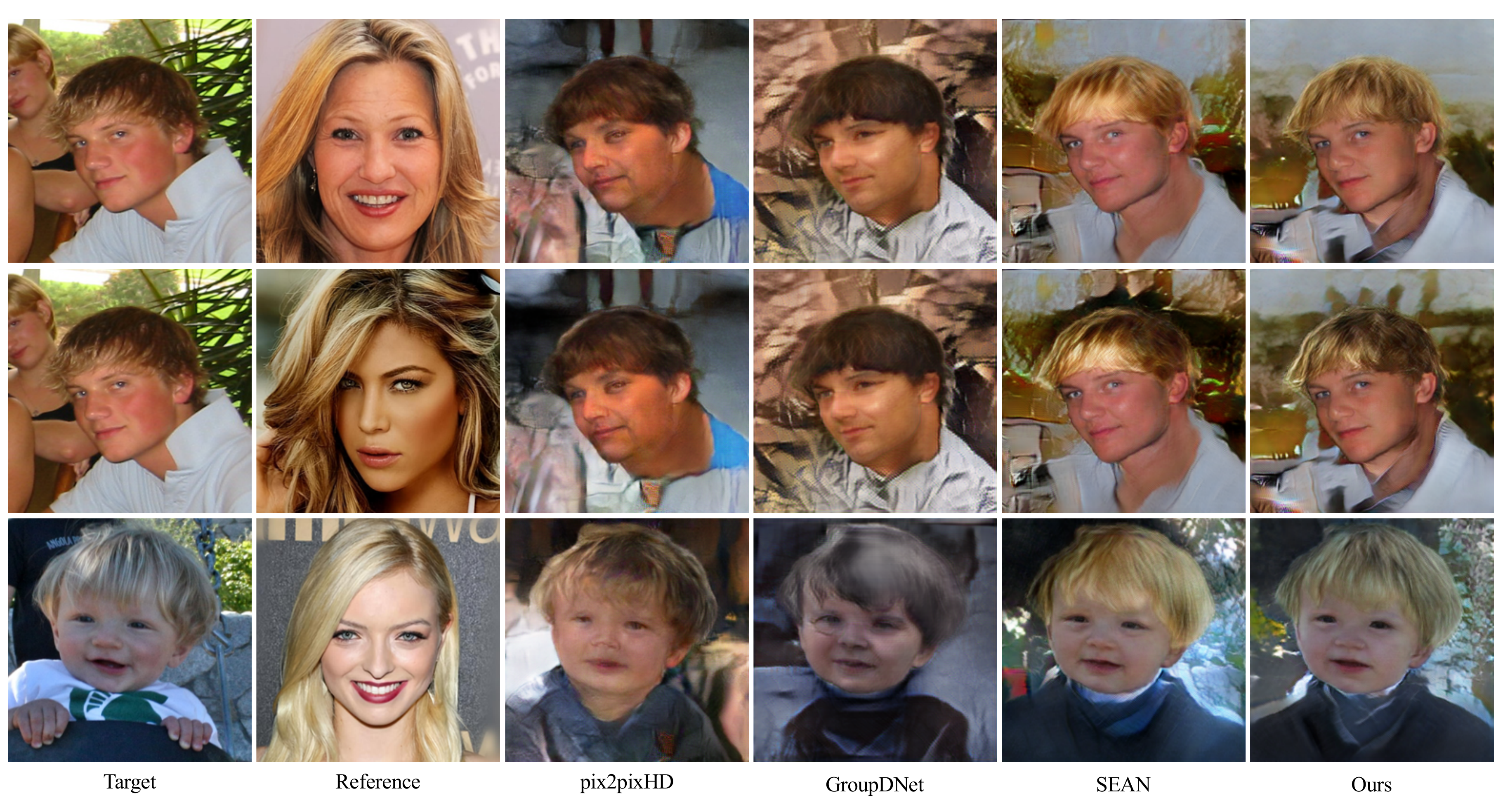}
\caption{Hair style transfer of our method and competing methods. }
\label{supp_harmony-hair}
\end{figure*}

\begin{figure*}[t]
\centering
\includegraphics[width=\linewidth]{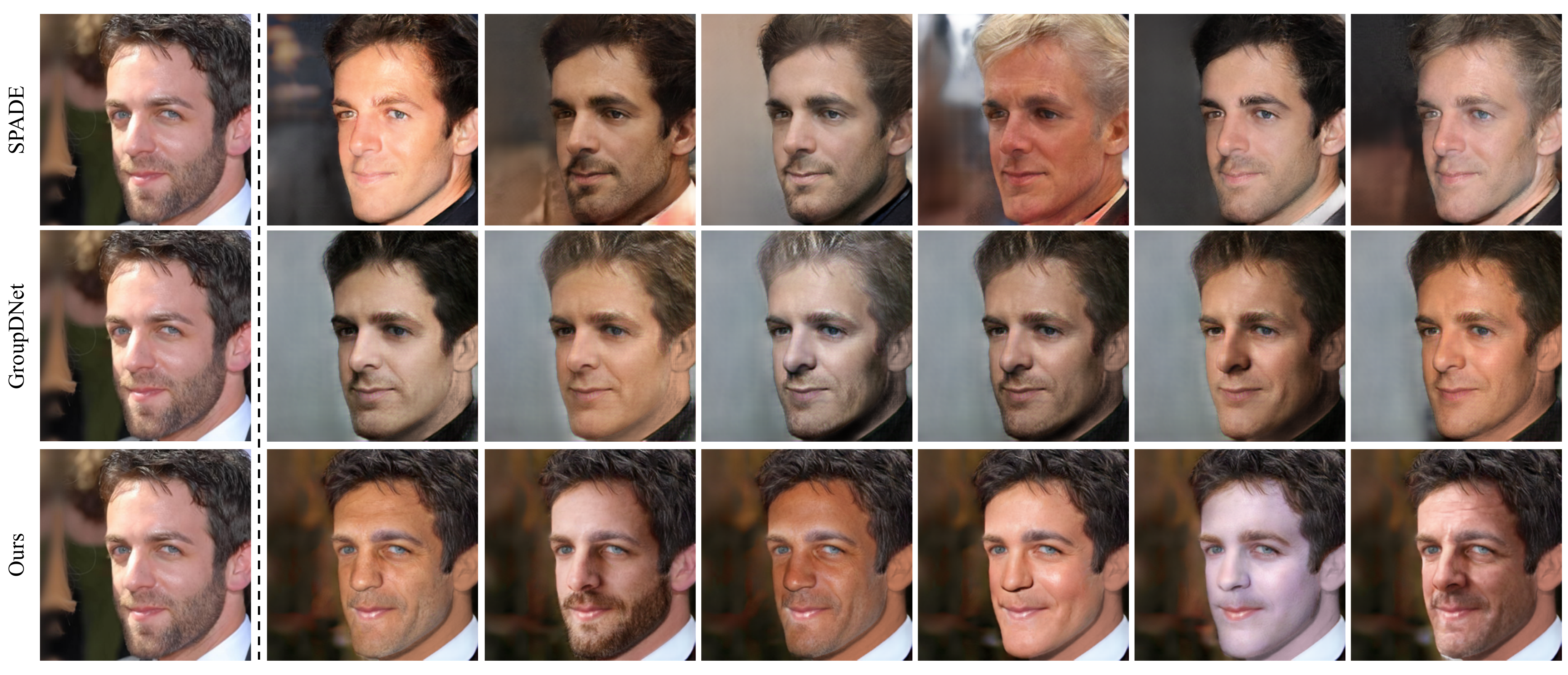}
\caption{Skin multi-modal synthesis. Our method achieves higher color and texture diversity of skin while preserving appearance of other regions, such as hair.}
\label{supp_modal-skin}
\end{figure*}

\begin{figure*}[t]
\centering
\includegraphics[width=\linewidth]{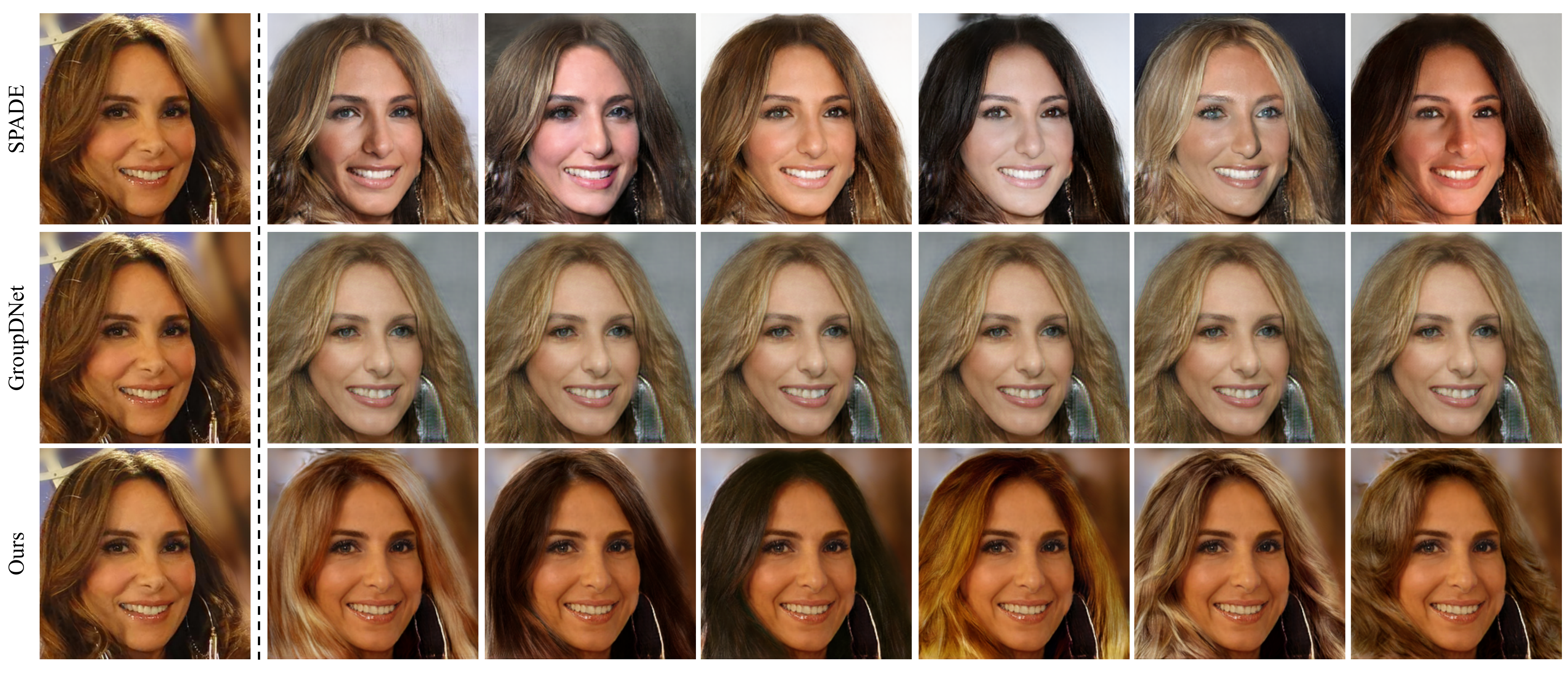}
\caption{Hair multi-modal synthesis. Our method achieves diverse hair textures, such as straight and curly hair. GroupDNet fails to generate diverse results for hair manipulation.}
\label{supp_modal-hair}
\end{figure*}

\paragraph{More results.}
We provide additional reference-guided face image synthesis and multi-modal manipulation results against state-of-the-art methods: pix2pixHD \cite{pix2pixhd}, GroupDNet \cite{groupdnet} and SEAN \cite{sean}. Figure \ref{supp_harmony-skin} and Figure \ref{supp_harmony-hair} show the results of skin and hair transfer, respectively. Our method achieves the best trade-off between image quality and regional transfer harmony. For hair transfer, we specificly transfer two references to the same target, as shown in the first two rows of Figure \ref{supp_harmony-hair}. GroupDNet generates the same results with different reference hair styles, and we can get the same conclusion from hair multi-modal synthesis(Figure \ref{supp_modal-hair}). Skin and hair multi-modal synthesis results are shown in Figure \ref{supp_modal-skin} and \ref{supp_modal-hair}, respectively. Our method achieves higher color and texture diversity while preserving appearance of the rest regions. 

%% file: Portrait_ST.bbl

\begin{thebibliography}{46}


\ifx \showCODEN    \undefined \def \showCODEN     #1{\unskip}     \fi
\ifx \showDOI      \undefined \def \showDOI       #1{#1}\fi
\ifx \showISBNx    \undefined \def \showISBNx     #1{\unskip}     \fi
\ifx \showISBNxiii \undefined \def \showISBNxiii  #1{\unskip}     \fi
\ifx \showISSN     \undefined \def \showISSN      #1{\unskip}     \fi
\ifx \showLCCN     \undefined \def \showLCCN      #1{\unskip}     \fi
\ifx \shownote     \undefined \def \shownote      #1{#1}          \fi
\ifx \showarticletitle \undefined \def \showarticletitle #1{#1}   \fi
\ifx \showURL      \undefined \def \showURL       {\relax}        \fi
\providecommand\bibfield[2]{#2}
\providecommand\bibinfo[2]{#2}
\providecommand\natexlab[1]{#1}
\providecommand\showeprint[2][]{arXiv:#2}

\bibitem[\protect\citeauthoryear{{Abdal}, {Zhu}, {Mitra}, and {Wonka}}{{Abdal}
  et~al\mbox{.}}{2020}]%
        {styleflow}
\bibfield{author}{\bibinfo{person}{Rameen {Abdal}}, \bibinfo{person}{Peihao
  {Zhu}}, \bibinfo{person}{Niloy~J. {Mitra}}, {and} \bibinfo{person}{Peter
  {Wonka}}.} \bibinfo{year}{2020}\natexlab{}.
\newblock \showarticletitle{StyleFlow: Attribute-conditioned Exploration of
  StyleGAN-Generated Images using Conditional Continuous Normalizing Flows.}
\newblock \bibinfo{journal}{\emph{arXiv preprint arXiv:2008.02401}}
  (\bibinfo{year}{2020}).
\newblock


\bibitem[\protect\citeauthoryear{{Arjovsky}, {Chintala}, and
  {Bottou}}{{Arjovsky} et~al\mbox{.}}{2017}]%
        {wgan}
\bibfield{author}{\bibinfo{person}{Martin {Arjovsky}}, \bibinfo{person}{Soumith
  {Chintala}}, {and} \bibinfo{person}{LÃ©on {Bottou}}.}
  \bibinfo{year}{2017}\natexlab{}.
\newblock \showarticletitle{Wasserstein Generative Adversarial Networks}. In
  \bibinfo{booktitle}{\emph{International Conference on Machine Learning}}.
  \bibinfo{pages}{214--223}.
\newblock


\bibitem[\protect\citeauthoryear{{Chen}, {Papandreou}, {Schroff}, and
  {Adam}}{{Chen} et~al\mbox{.}}{2017}]%
        {deeplabv3}
\bibfield{author}{\bibinfo{person}{Liang-Chieh {Chen}}, \bibinfo{person}{George
  {Papandreou}}, \bibinfo{person}{Florian {Schroff}}, {and}
  \bibinfo{person}{Hartwig {Adam}}.} \bibinfo{year}{2017}\natexlab{}.
\newblock \showarticletitle{Rethinking Atrous Convolution for Semantic Image
  Segmentation}.
\newblock \bibinfo{journal}{\emph{arXiv preprint arXiv:1706.05587}}
  (\bibinfo{year}{2017}).
\newblock


\bibitem[\protect\citeauthoryear{Chen and Koltun}{Chen and Koltun}{2017}]%
        {crn}
\bibfield{author}{\bibinfo{person}{Qifeng Chen} {and} \bibinfo{person}{Vladlen
  Koltun}.} \bibinfo{year}{2017}\natexlab{}.
\newblock \showarticletitle{Photographic Image Synthesis with Cascaded
  Refinement Networks}. In \bibinfo{booktitle}{\emph{2017 IEEE International
  Conference on Computer Vision (ICCV)}}.
\newblock


\bibitem[\protect\citeauthoryear{{Choi}, {Uh}, {Yoo}, and {Ha}}{{Choi}
  et~al\mbox{.}}{2020}]%
        {starganv2}
\bibfield{author}{\bibinfo{person}{Yunjey {Choi}}, \bibinfo{person}{Youngjung
  {Uh}}, \bibinfo{person}{Jaejun {Yoo}}, {and} \bibinfo{person}{Jung-Woo
  {Ha}}.} \bibinfo{year}{2020}\natexlab{}.
\newblock \showarticletitle{StarGAN v2: Diverse Image Synthesis for Multiple
  Domains}. In \bibinfo{booktitle}{\emph{2020 IEEE/CVF Conference on Computer
  Vision and Pattern Recognition (CVPR)}}. \bibinfo{pages}{8188--8197}.
\newblock


\bibitem[\protect\citeauthoryear{{Cong}, {Zhang}, {Niu}, {Liu}, {Ling}, {Li},
  and {Zhang}}{{Cong} et~al\mbox{.}}{2020}]%
        {dovenet}
\bibfield{author}{\bibinfo{person}{Wenyan {Cong}}, \bibinfo{person}{Jianfu
  {Zhang}}, \bibinfo{person}{Li {Niu}}, \bibinfo{person}{Liu {Liu}},
  \bibinfo{person}{Zhixin {Ling}}, \bibinfo{person}{Weiyuan {Li}}, {and}
  \bibinfo{person}{Liqing {Zhang}}.} \bibinfo{year}{2020}\natexlab{}.
\newblock \showarticletitle{DoveNet: Deep Image Harmonization via Domain
  Verification}. In \bibinfo{booktitle}{\emph{2020 IEEE/CVF Conference on
  Computer Vision and Pattern Recognition (CVPR)}}.
  \bibinfo{pages}{8394--8403}.
\newblock


\bibitem[\protect\citeauthoryear{{Cun} and {Pun}}{{Cun} and {Pun}}{2020}]%
        {tip_harmony}
\bibfield{author}{\bibinfo{person}{Xiaodong {Cun}} {and}
  \bibinfo{person}{Chi-Man {Pun}}.} \bibinfo{year}{2020}\natexlab{}.
\newblock \showarticletitle{Improving the Harmony of the Composite Image by
  Spatial-Separated Attention Module}.
\newblock \bibinfo{journal}{\emph{IEEE Transactions on Image Processing}}
  \bibinfo{volume}{29} (\bibinfo{year}{2020}), \bibinfo{pages}{4759--4771}.
\newblock


\bibitem[\protect\citeauthoryear{{Denton}, {Chintala}, {Szlam}, and
  {Fergus}}{{Denton} et~al\mbox{.}}{2015}]%
        {laplacian-gan}
\bibfield{author}{\bibinfo{person}{Emily {Denton}}, \bibinfo{person}{Soumith
  {Chintala}}, \bibinfo{person}{Arthur {Szlam}}, {and} \bibinfo{person}{Rob
  {Fergus}}.} \bibinfo{year}{2015}\natexlab{}.
\newblock \showarticletitle{Deep generative image models using a Laplacian
  pyramid of adversarial networks}. In \bibinfo{booktitle}{\emph{NIPS'15
  Proceedings of the 28th International Conference on Neural Information
  Processing Systems - Volume 1}}, Vol.~\bibinfo{volume}{28}.
  \bibinfo{pages}{1486--1494}.
\newblock


\bibitem[\protect\citeauthoryear{Goodfellow, Pouget-Abadie, Mirza, Xu,
  Warde-Farley, Ozair, Courville, and Bengio}{Goodfellow et~al\mbox{.}}{2014}]%
        {gan}
\bibfield{author}{\bibinfo{person}{Ian~J. Goodfellow}, \bibinfo{person}{Jean
  Pouget-Abadie}, \bibinfo{person}{Mehdi Mirza}, \bibinfo{person}{Bing Xu},
  \bibinfo{person}{David Warde-Farley}, \bibinfo{person}{Sherjil Ozair},
  \bibinfo{person}{Aaron Courville}, {and} \bibinfo{person}{Yoshua Bengio}.}
  \bibinfo{year}{2014}\natexlab{}.
\newblock \showarticletitle{Generative Adversarial Networks}.
\newblock \bibinfo{journal}{\emph{Advances in Neural Information Processing
  Systems}}  \bibinfo{volume}{3} (\bibinfo{year}{2014}),
  \bibinfo{pages}{2672--2680}.
\newblock


\bibitem[\protect\citeauthoryear{Gu, Bao, Yang, Chen, Wen, and Yuan}{Gu
  et~al\mbox{.}}{2019}]%
        {mask2face}
\bibfield{author}{\bibinfo{person}{Shuyang Gu}, \bibinfo{person}{Jianmin Bao},
  \bibinfo{person}{Hao Yang}, \bibinfo{person}{Dong Chen},
  \bibinfo{person}{Fang Wen}, {and} \bibinfo{person}{Lu Yuan}.}
  \bibinfo{year}{2019}\natexlab{}.
\newblock \showarticletitle{Mask-Guided Portrait Editing With Conditional
  GANs}. In \bibinfo{booktitle}{\emph{Proceedings of the IEEE Conference on
  Computer Vision and Pattern Recognition}}. \bibinfo{pages}{3436--3445}.
\newblock


\bibitem[\protect\citeauthoryear{{Gulrajani}, {Ahmed}, {Arjovsky}, {Dumoulin},
  and {Courville}}{{Gulrajani} et~al\mbox{.}}{2017}]%
        {gp-wgan}
\bibfield{author}{\bibinfo{person}{Ishaan {Gulrajani}}, \bibinfo{person}{Faruk
  {Ahmed}}, \bibinfo{person}{Martin {Arjovsky}}, \bibinfo{person}{Vincent
  {Dumoulin}}, {and} \bibinfo{person}{Aaron {Courville}}.}
  \bibinfo{year}{2017}\natexlab{}.
\newblock \showarticletitle{Improved training of wasserstein GANs}. In
  \bibinfo{booktitle}{\emph{NIPS'17 Proceedings of the 31st International
  Conference on Neural Information Processing Systems}},
  Vol.~\bibinfo{volume}{30}. \bibinfo{pages}{5769--5779}.
\newblock


\bibitem[\protect\citeauthoryear{{Heusel}, {Ramsauer}, {Unterthiner},
  {Nessler}, and {Hochreiter}}{{Heusel} et~al\mbox{.}}{2017}]%
        {ttur}
\bibfield{author}{\bibinfo{person}{Martin {Heusel}}, \bibinfo{person}{Hubert
  {Ramsauer}}, \bibinfo{person}{Thomas {Unterthiner}},
  \bibinfo{person}{Bernhard {Nessler}}, {and} \bibinfo{person}{Sepp
  {Hochreiter}}.} \bibinfo{year}{2017}\natexlab{}.
\newblock \showarticletitle{GANs Trained by a Two Time-Scale Update Rule
  Converge to a Local Nash Equilibrium}. In \bibinfo{booktitle}{\emph{Advances
  in Neural Information Processing Systems}}, Vol.~\bibinfo{volume}{30}.
  \bibinfo{pages}{6626--6637}.
\newblock


\bibitem[\protect\citeauthoryear{Huang, Liu, Belongie, and Kautz}{Huang
  et~al\mbox{.}}{2018}]%
        {munit}
\bibfield{author}{\bibinfo{person}{Xun Huang}, \bibinfo{person}{Ming-Yu Liu},
  \bibinfo{person}{Serge Belongie}, {and} \bibinfo{person}{Jan Kautz}.}
  \bibinfo{year}{2018}\natexlab{}.
\newblock \showarticletitle{Multimodal Unsupervised Image-to-image
  Translation}. In \bibinfo{booktitle}{\emph{ECCV}}.
\newblock


\bibitem[\protect\citeauthoryear{Isola, Zhu, Zhou, and Efros}{Isola
  et~al\mbox{.}}{2017}]%
        {pix2pix}
\bibfield{author}{\bibinfo{person}{Phillip Isola}, \bibinfo{person}{Jun-Yan
  Zhu}, \bibinfo{person}{Tinghui Zhou}, {and} \bibinfo{person}{Alexei~A
  Efros}.} \bibinfo{year}{2017}\natexlab{}.
\newblock \showarticletitle{Image-to-Image Translation with Conditional
  Adversarial Networks}. In \bibinfo{booktitle}{\emph{Computer Vision and
  Pattern Recognition (CVPR), 2017 IEEE Conference on}}.
\newblock


\bibitem[\protect\citeauthoryear{{Jiang}, {Liu}, {Gao}, {Cao}, {He}, {Feng},
  and {Yan}}{{Jiang} et~al\mbox{.}}{2020}]%
        {psgan}
\bibfield{author}{\bibinfo{person}{Wentao {Jiang}}, \bibinfo{person}{Si {Liu}},
  \bibinfo{person}{Chen {Gao}}, \bibinfo{person}{Jie {Cao}},
  \bibinfo{person}{Ran {He}}, \bibinfo{person}{Jiashi {Feng}}, {and}
  \bibinfo{person}{Shuicheng {Yan}}.} \bibinfo{year}{2020}\natexlab{}.
\newblock \showarticletitle{PSGAN: Pose and Expression Robust Spatial-Aware GAN
  for Customizable Makeup Transfer}. In \bibinfo{booktitle}{\emph{2020 IEEE/CVF
  Conference on Computer Vision and Pattern Recognition (CVPR)}}.
  \bibinfo{pages}{5194--5202}.
\newblock


\bibitem[\protect\citeauthoryear{Karras, Aila, Laine, and Lehtinen}{Karras
  et~al\mbox{.}}{2018}]%
        {pggan}
\bibfield{author}{\bibinfo{person}{Tero Karras}, \bibinfo{person}{Timo Aila},
  \bibinfo{person}{Samuli Laine}, {and} \bibinfo{person}{Jaakko Lehtinen}.}
  \bibinfo{year}{2018}\natexlab{}.
\newblock \showarticletitle{Progressive Growing of GANs for Improved Quality,
  Stability, and Variation}.
\newblock \bibinfo{journal}{\emph{international conference on learning
  representations}} (\bibinfo{year}{2018}).
\newblock


\bibitem[\protect\citeauthoryear{Karras, Laine, and Aila}{Karras
  et~al\mbox{.}}{2019}]%
        {stylegan}
\bibfield{author}{\bibinfo{person}{Tero Karras}, \bibinfo{person}{Samuli
  Laine}, {and} \bibinfo{person}{Timo Aila}.} \bibinfo{year}{2019}\natexlab{}.
\newblock \showarticletitle{A Style-Based Generator Architecture for Generative
  Adversarial Networks}. In \bibinfo{booktitle}{\emph{Proceedings of the
  IEEE/CVF Conference on Computer Vision and Pattern Recognition (CVPR)}}.
\newblock


\bibitem[\protect\citeauthoryear{Karras, Laine, Aittala, Hellsten, Lehtinen,
  and Aila}{Karras et~al\mbox{.}}{2020}]%
        {stylegan2}
\bibfield{author}{\bibinfo{person}{Tero Karras}, \bibinfo{person}{Samuli
  Laine}, \bibinfo{person}{Miika Aittala}, \bibinfo{person}{Janne Hellsten},
  \bibinfo{person}{Jaakko Lehtinen}, {and} \bibinfo{person}{Timo Aila}.}
  \bibinfo{year}{2020}\natexlab{}.
\newblock \showarticletitle{Analyzing and Improving the Image Quality of
  StyleGAN}. In \bibinfo{booktitle}{\emph{Proceedings of the IEEE/CVF
  Conference on Computer Vision and Pattern Recognition (CVPR)}}.
\newblock


\bibitem[\protect\citeauthoryear{{Kingma} and {Ba}}{{Kingma} and {Ba}}{2015}]%
        {adam}
\bibfield{author}{\bibinfo{person}{Diederik~P. {Kingma}} {and}
  \bibinfo{person}{Jimmy~Lei {Ba}}.} \bibinfo{year}{2015}\natexlab{}.
\newblock \showarticletitle{Adam: A Method for Stochastic Optimization}. In
  \bibinfo{booktitle}{\emph{ICLR 2015 : International Conference on Learning
  Representations 2015}}.
\newblock


\bibitem[\protect\citeauthoryear{{Kingma} and {Welling}}{{Kingma} and
  {Welling}}{2014}]%
        {vae}
\bibfield{author}{\bibinfo{person}{Diederik~P {Kingma}} {and}
  \bibinfo{person}{Max {Welling}}.} \bibinfo{year}{2014}\natexlab{}.
\newblock \showarticletitle{Auto-Encoding Variational Bayes}. In
  \bibinfo{booktitle}{\emph{ICLR 2014 : International Conference on Learning
  Representations (ICLR) 2014}}.
\newblock


\bibitem[\protect\citeauthoryear{{Lee}, {Liu}, {Wu}, and {Luo}}{{Lee}
  et~al\mbox{.}}{2020b}]%
        {maskgan}
\bibfield{author}{\bibinfo{person}{Cheng-Han {Lee}}, \bibinfo{person}{Ziwei
  {Liu}}, \bibinfo{person}{Lingyun {Wu}}, {and} \bibinfo{person}{Ping {Luo}}.}
  \bibinfo{year}{2020}\natexlab{b}.
\newblock \showarticletitle{MaskGAN: Towards Diverse and Interactive Facial
  Image Manipulation}. In \bibinfo{booktitle}{\emph{2020 IEEE/CVF Conference on
  Computer Vision and Pattern Recognition (CVPR)}}.
  \bibinfo{pages}{5549--5558}.
\newblock


\bibitem[\protect\citeauthoryear{Lee, Tseng, Huang, Singh, and Yang}{Lee
  et~al\mbox{.}}{2018}]%
        {drit}
\bibfield{author}{\bibinfo{person}{Hsin-Ying Lee}, \bibinfo{person}{Hung-Yu
  Tseng}, \bibinfo{person}{Jia-Bin Huang}, \bibinfo{person}{Maneesh~Kumar
  Singh}, {and} \bibinfo{person}{Ming-Hsuan Yang}.}
  \bibinfo{year}{2018}\natexlab{}.
\newblock \showarticletitle{Diverse Image-to-Image Translation via Disentangled
  Representations}. In \bibinfo{booktitle}{\emph{European Conference on
  Computer Vision}}.
\newblock


\bibitem[\protect\citeauthoryear{{Lee}, {Kim}, {Lee}, {Kim}, {Chang}, and
  {Choo}}{{Lee} et~al\mbox{.}}{2020a}]%
        {refsketch}
\bibfield{author}{\bibinfo{person}{Junsoo {Lee}}, \bibinfo{person}{Eungyeup
  {Kim}}, \bibinfo{person}{Yunsung {Lee}}, \bibinfo{person}{Dongjun {Kim}},
  \bibinfo{person}{Jaehyuk {Chang}}, {and} \bibinfo{person}{Jaegul {Choo}}.}
  \bibinfo{year}{2020}\natexlab{a}.
\newblock \showarticletitle{Reference-Based Sketch Image Colorization Using
  Augmented-Self Reference and Dense Semantic Correspondence}. In
  \bibinfo{booktitle}{\emph{2020 IEEE/CVF Conference on Computer Vision and
  Pattern Recognition (CVPR)}}. \bibinfo{pages}{5801--5810}.
\newblock


\bibitem[\protect\citeauthoryear{Liu, Shi, Shen, Si, Wang, and Mei}{Liu
  et~al\mbox{.}}{2020}]%
        {lapa}
\bibfield{author}{\bibinfo{person}{Yinglu Liu}, \bibinfo{person}{Hailin Shi},
  \bibinfo{person}{Hao Shen}, \bibinfo{person}{Yue Si}, \bibinfo{person}{Xiaobo
  Wang}, {and} \bibinfo{person}{Tao Mei}.} \bibinfo{year}{2020}\natexlab{}.
\newblock \showarticletitle{A New Dataset and Boundary-Attention Semantic
  Segmentation for Face Parsing.}. In \bibinfo{booktitle}{\emph{AAAI}}.
  \bibinfo{pages}{11637--11644}.
\newblock


\bibitem[\protect\citeauthoryear{{Mao}, {Li}, {Xie}, {Lau}, {Wang}, and
  {Smolley}}{{Mao} et~al\mbox{.}}{2017}]%
        {lsgan}
\bibfield{author}{\bibinfo{person}{Xudong {Mao}}, \bibinfo{person}{Qing {Li}},
  \bibinfo{person}{Haoran {Xie}}, \bibinfo{person}{Raymond~Y.K. {Lau}},
  \bibinfo{person}{Zhen {Wang}}, {and} \bibinfo{person}{Stephen~Paul
  {Smolley}}.} \bibinfo{year}{2017}\natexlab{}.
\newblock \showarticletitle{Least Squares Generative Adversarial Networks}. In
  \bibinfo{booktitle}{\emph{2017 IEEE International Conference on Computer
  Vision (ICCV)}}. \bibinfo{pages}{2813--2821}.
\newblock


\bibitem[\protect\citeauthoryear{{Miyato}, {Kataoka}, {Koyama}, and
  {Yoshida}}{{Miyato} et~al\mbox{.}}{2018}]%
        {sn}
\bibfield{author}{\bibinfo{person}{Takeru {Miyato}}, \bibinfo{person}{Toshiki
  {Kataoka}}, \bibinfo{person}{Masanori {Koyama}}, {and}
  \bibinfo{person}{Yuichi {Yoshida}}.} \bibinfo{year}{2018}\natexlab{}.
\newblock \showarticletitle{Spectral Normalization for Generative Adversarial
  Networks}. In \bibinfo{booktitle}{\emph{International Conference on Learning
  Representations}}.
\newblock


\bibitem[\protect\citeauthoryear{{Park}, {Liu}, {Wang}, and {Zhu}}{{Park}
  et~al\mbox{.}}{2019}]%
        {spade}
\bibfield{author}{\bibinfo{person}{Taesung {Park}}, \bibinfo{person}{Ming-Yu
  {Liu}}, \bibinfo{person}{Ting-Chun {Wang}}, {and} \bibinfo{person}{Jun-Yan
  {Zhu}}.} \bibinfo{year}{2019}\natexlab{}.
\newblock \showarticletitle{Semantic Image Synthesis With Spatially-Adaptive
  Normalization}. In \bibinfo{booktitle}{\emph{2019 IEEE/CVF Conference on
  Computer Vision and Pattern Recognition (CVPR)}}.
  \bibinfo{pages}{2337--2346}.
\newblock


\bibitem[\protect\citeauthoryear{{Rezende} and {Mohamed}}{{Rezende} and
  {Mohamed}}{2015}]%
        {nf}
\bibfield{author}{\bibinfo{person}{Danilo {Rezende}} {and}
  \bibinfo{person}{Shakir {Mohamed}}.} \bibinfo{year}{2015}\natexlab{}.
\newblock \showarticletitle{Variational Inference with Normalizing Flows}. In
  \bibinfo{booktitle}{\emph{Proceedings of The 32nd International Conference on
  Machine Learning}}. \bibinfo{pages}{1530--1538}.
\newblock


\bibitem[\protect\citeauthoryear{{Richardson}, {Alaluf}, {Patashnik}, {Nitzan},
  {Azar}, {Shapiro}, and {Cohen-or}}{{Richardson} et~al\mbox{.}}{2021}]%
        {encoding_stylegan}
\bibfield{author}{\bibinfo{person}{Elad {Richardson}}, \bibinfo{person}{Yuval
  {Alaluf}}, \bibinfo{person}{Or {Patashnik}}, \bibinfo{person}{Yotam
  {Nitzan}}, \bibinfo{person}{Yaniv {Azar}}, \bibinfo{person}{Stav {Shapiro}},
  {and} \bibinfo{person}{Daniel {Cohen-or}}.} \bibinfo{year}{2021}\natexlab{}.
\newblock \showarticletitle{Encoding in Style: a StyleGAN Encoder for
  Image-to-Image Translation}. In \bibinfo{booktitle}{\emph{arxiv:cs.CV}}.
\newblock


\bibitem[\protect\citeauthoryear{{Shen}, {Gu}, {Tang}, and {Zhou}}{{Shen}
  et~al\mbox{.}}{2020}]%
        {interfacegan}
\bibfield{author}{\bibinfo{person}{Yujun {Shen}}, \bibinfo{person}{Jinjin
  {Gu}}, \bibinfo{person}{Xiaoou {Tang}}, {and} \bibinfo{person}{Bolei
  {Zhou}}.} \bibinfo{year}{2020}\natexlab{}.
\newblock \showarticletitle{Interpreting the Latent Space of GANs for Semantic
  Face Editing}. In \bibinfo{booktitle}{\emph{2020 IEEE/CVF Conference on
  Computer Vision and Pattern Recognition (CVPR)}}.
  \bibinfo{pages}{9243--9252}.
\newblock


\bibitem[\protect\citeauthoryear{{Shen} and {Zhou}}{{Shen} and {Zhou}}{2020}]%
        {sefa}
\bibfield{author}{\bibinfo{person}{Yujun {Shen}} {and} \bibinfo{person}{Bolei
  {Zhou}}.} \bibinfo{year}{2020}\natexlab{}.
\newblock \showarticletitle{Closed-Form Factorization of Latent Semantics in
  GANs}.
\newblock \bibinfo{journal}{\emph{arXiv preprint arXiv:2007.06600}}
  (\bibinfo{year}{2020}).
\newblock


\bibitem[\protect\citeauthoryear{{Tewari}, {Elgharib}, {Bharaj}, {Bernard},
  {Seidel}, {Perez}, {Zollhofer}, and {Theobalt}}{{Tewari}
  et~al\mbox{.}}{2020}]%
        {stylerig}
\bibfield{author}{\bibinfo{person}{Ayush {Tewari}}, \bibinfo{person}{Mohamed
  {Elgharib}}, \bibinfo{person}{Gaurav {Bharaj}}, \bibinfo{person}{Florian
  {Bernard}}, \bibinfo{person}{Hans-Peter {Seidel}}, \bibinfo{person}{Patrick
  {Perez}}, \bibinfo{person}{Michael {Zollhofer}}, {and}
  \bibinfo{person}{Christian {Theobalt}}.} \bibinfo{year}{2020}\natexlab{}.
\newblock \showarticletitle{StyleRig: Rigging StyleGAN for 3D Control Over
  Portrait Images}. In \bibinfo{booktitle}{\emph{2020 IEEE/CVF Conference on
  Computer Vision and Pattern Recognition (CVPR)}}.
  \bibinfo{pages}{6142--6151}.
\newblock


\bibitem[\protect\citeauthoryear{{Tsai}, {Shen}, {Lin}, {Sunkavalli}, {Lu}, and
  {Yang}}{{Tsai} et~al\mbox{.}}{2017}]%
        {dih}
\bibfield{author}{\bibinfo{person}{Yi-Hsuan {Tsai}}, \bibinfo{person}{Xiaohui
  {Shen}}, \bibinfo{person}{Zhe {Lin}}, \bibinfo{person}{Kalyan {Sunkavalli}},
  \bibinfo{person}{Xin {Lu}}, {and} \bibinfo{person}{Ming-Hsuan {Yang}}.}
  \bibinfo{year}{2017}\natexlab{}.
\newblock \showarticletitle{Deep Image Harmonization}. In
  \bibinfo{booktitle}{\emph{2017 IEEE Conference on Computer Vision and Pattern
  Recognition (CVPR)}}. \bibinfo{pages}{2799--2807}.
\newblock


\bibitem[\protect\citeauthoryear{{Vaswani}, {Shazeer}, {Parmar}, {Uszkoreit},
  {Jones}, {Gomez}, {Kaiser}, and {Polosukhin}}{{Vaswani}
  et~al\mbox{.}}{2017}]%
        {transformer}
\bibfield{author}{\bibinfo{person}{Ashish {Vaswani}}, \bibinfo{person}{Noam
  {Shazeer}}, \bibinfo{person}{Niki {Parmar}}, \bibinfo{person}{Jakob
  {Uszkoreit}}, \bibinfo{person}{Llion {Jones}}, \bibinfo{person}{Aidan~N.
  {Gomez}}, \bibinfo{person}{Lukasz {Kaiser}}, {and} \bibinfo{person}{Illia
  {Polosukhin}}.} \bibinfo{year}{2017}\natexlab{}.
\newblock \showarticletitle{Attention is All You Need}. In
  \bibinfo{booktitle}{\emph{Proceedings of the 31st International Conference on
  Neural Information Processing Systems}}, Vol.~\bibinfo{volume}{30}.
  \bibinfo{pages}{5998--6008}.
\newblock


\bibitem[\protect\citeauthoryear{{Wang}, {Yang}, {Li}, {Liang}, {Zhang},
  {Hall}, and {Hu}}{{Wang} et~al\mbox{.}}{2019}]%
        {exampleguide}
\bibfield{author}{\bibinfo{person}{Miao {Wang}}, \bibinfo{person}{Guo-Ye
  {Yang}}, \bibinfo{person}{Ruilong {Li}}, \bibinfo{person}{Run-Ze {Liang}},
  \bibinfo{person}{Song-Hai {Zhang}}, \bibinfo{person}{Peter~M. {Hall}}, {and}
  \bibinfo{person}{Shi-Min {Hu}}.} \bibinfo{year}{2019}\natexlab{}.
\newblock \showarticletitle{Example-Guided Style-Consistent Image Synthesis
  From Semantic Labeling}. In \bibinfo{booktitle}{\emph{2019 IEEE/CVF
  Conference on Computer Vision and Pattern Recognition (CVPR)}}.
  \bibinfo{pages}{1495--1504}.
\newblock


\bibitem[\protect\citeauthoryear{{Wang}, {Liu}, {Zhu}, {Tao}, {Kautz}, and
  {Catanzaro}}{{Wang} et~al\mbox{.}}{2018b}]%
        {pix2pixhd}
\bibfield{author}{\bibinfo{person}{Ting-Chun {Wang}}, \bibinfo{person}{Ming-Yu
  {Liu}}, \bibinfo{person}{Jun-Yan {Zhu}}, \bibinfo{person}{Andrew {Tao}},
  \bibinfo{person}{Jan {Kautz}}, {and} \bibinfo{person}{Bryan {Catanzaro}}.}
  \bibinfo{year}{2018}\natexlab{b}.
\newblock \showarticletitle{High-Resolution Image Synthesis and Semantic
  Manipulation with Conditional GANs}. In \bibinfo{booktitle}{\emph{2018
  IEEE/CVF Conference on Computer Vision and Pattern Recognition}}.
  \bibinfo{pages}{8798--8807}.
\newblock


\bibitem[\protect\citeauthoryear{{Wang}, {Girshick}, {Gupta}, and {He}}{{Wang}
  et~al\mbox{.}}{2018a}]%
        {nonlocal}
\bibfield{author}{\bibinfo{person}{Xiaolong {Wang}}, \bibinfo{person}{Ross
  {Girshick}}, \bibinfo{person}{Abhinav {Gupta}}, {and}
  \bibinfo{person}{Kaiming {He}}.} \bibinfo{year}{2018}\natexlab{a}.
\newblock \showarticletitle{Non-local Neural Networks}. In
  \bibinfo{booktitle}{\emph{2018 IEEE/CVF Conference on Computer Vision and
  Pattern Recognition}}. \bibinfo{pages}{7794--7803}.
\newblock


\bibitem[\protect\citeauthoryear{{Yang}, {Hong}, {Jang}, {Zhao}, and
  {Lee}}{{Yang} et~al\mbox{.}}{2019}]%
        {dsgan}
\bibfield{author}{\bibinfo{person}{Dingdong {Yang}}, \bibinfo{person}{Seunghoon
  {Hong}}, \bibinfo{person}{Yunseok {Jang}}, \bibinfo{person}{Tianchen {Zhao}},
  {and} \bibinfo{person}{Honglak {Lee}}.} \bibinfo{year}{2019}\natexlab{}.
\newblock \showarticletitle{Diversity-Sensitive Conditional Generative
  Adversarial Networks}. In \bibinfo{booktitle}{\emph{International Conference
  on Learning Representations}}.
\newblock


\bibitem[\protect\citeauthoryear{{Zhang}, {Goodfellow}, {Metaxas}, and
  {Odena}}{{Zhang} et~al\mbox{.}}{2018}]%
        {sagan}
\bibfield{author}{\bibinfo{person}{Han {Zhang}}, \bibinfo{person}{Ian~J.
  {Goodfellow}}, \bibinfo{person}{Dimitris~N. {Metaxas}}, {and}
  \bibinfo{person}{Augustus {Odena}}.} \bibinfo{year}{2018}\natexlab{}.
\newblock \showarticletitle{Self-Attention Generative Adversarial Networks}. In
  \bibinfo{booktitle}{\emph{International Conference on Machine Learning}}.
  \bibinfo{pages}{7354--7363}.
\newblock


\bibitem[\protect\citeauthoryear{{Zhang}, {Zhang}, {Chen}, {Yuan}, and
  {Wen}}{{Zhang} et~al\mbox{.}}{2020}]%
        {cocosnet}
\bibfield{author}{\bibinfo{person}{Pan {Zhang}}, \bibinfo{person}{Bo {Zhang}},
  \bibinfo{person}{Dong {Chen}}, \bibinfo{person}{Lu {Yuan}}, {and}
  \bibinfo{person}{Fang {Wen}}.} \bibinfo{year}{2020}\natexlab{}.
\newblock \showarticletitle{Cross-Domain Correspondence Learning for
  Exemplar-Based Image Translation}. In \bibinfo{booktitle}{\emph{2020 IEEE/CVF
  Conference on Computer Vision and Pattern Recognition (CVPR)}}.
  \bibinfo{pages}{5143--5153}.
\newblock


\bibitem[\protect\citeauthoryear{Zhang, Isola, Efros, Shechtman, and
  Wang}{Zhang et~al\mbox{.}}{2018}]%
        {lpips}
\bibfield{author}{\bibinfo{person}{Richard Zhang}, \bibinfo{person}{Phillip
  Isola}, \bibinfo{person}{Alexei~A Efros}, \bibinfo{person}{Eli Shechtman},
  {and} \bibinfo{person}{Oliver Wang}.} \bibinfo{year}{2018}\natexlab{}.
\newblock \showarticletitle{The Unreasonable Effectiveness of Deep Features as
  a Perceptual Metric}. In \bibinfo{booktitle}{\emph{CVPR}}.
\newblock


\bibitem[\protect\citeauthoryear{Zhu, Shen, Zhao, and Zhou}{Zhu
  et~al\mbox{.}}{2020}]%
        {idinvert}
\bibfield{author}{\bibinfo{person}{Jiapeng Zhu}, \bibinfo{person}{Yujun Shen},
  \bibinfo{person}{Deli Zhao}, {and} \bibinfo{person}{Bolei Zhou}.}
  \bibinfo{year}{2020}\natexlab{}.
\newblock \showarticletitle{In-domain GAN Inversion for Real Image Editing}. In
  \bibinfo{booktitle}{\emph{Proceedings of European Conference on Computer
  Vision (ECCV)}}.
\newblock


\bibitem[\protect\citeauthoryear{{Zhu}, {Krahenbuhl}, {Shechtman}, and
  {Efros}}{{Zhu} et~al\mbox{.}}{2015}]%
        {zhu_harmony}
\bibfield{author}{\bibinfo{person}{Jun-Yan {Zhu}}, \bibinfo{person}{Philipp
  {Krahenbuhl}}, \bibinfo{person}{Eli {Shechtman}}, {and}
  \bibinfo{person}{Alexei~A. {Efros}}.} \bibinfo{year}{2015}\natexlab{}.
\newblock \showarticletitle{Learning a Discriminative Model for the Perception
  of Realism in Composite Images}. In \bibinfo{booktitle}{\emph{2015 IEEE
  International Conference on Computer Vision (ICCV)}}.
  \bibinfo{pages}{3943--3951}.
\newblock


\bibitem[\protect\citeauthoryear{Zhu, Zhang, Pathak, Darrell, Efros, Wang, and
  Shechtman}{Zhu et~al\mbox{.}}{2017}]%
        {bicyclegan}
\bibfield{author}{\bibinfo{person}{Jun-Yan Zhu}, \bibinfo{person}{Richard
  Zhang}, \bibinfo{person}{Deepak Pathak}, \bibinfo{person}{Trevor Darrell},
  \bibinfo{person}{Alexei~A Efros}, \bibinfo{person}{Oliver Wang}, {and}
  \bibinfo{person}{Eli Shechtman}.} \bibinfo{year}{2017}\natexlab{}.
\newblock \showarticletitle{Toward multimodal image-to-image translation}. In
  \bibinfo{booktitle}{\emph{Advances in Neural Information Processing
  Systems}}.
\newblock


\bibitem[\protect\citeauthoryear{{Zhu}, {Abdal}, {Qin}, and {Wonka}}{{Zhu}
  et~al\mbox{.}}{2020}]%
        {sean}
\bibfield{author}{\bibinfo{person}{Peihao {Zhu}}, \bibinfo{person}{Rameen
  {Abdal}}, \bibinfo{person}{Yipeng {Qin}}, {and} \bibinfo{person}{Peter
  {Wonka}}.} \bibinfo{year}{2020}\natexlab{}.
\newblock \showarticletitle{SEAN: Image Synthesis With Semantic Region-Adaptive
  Normalization}. In \bibinfo{booktitle}{\emph{2020 IEEE/CVF Conference on
  Computer Vision and Pattern Recognition (CVPR)}}.
  \bibinfo{pages}{5104--5113}.
\newblock


\bibitem[\protect\citeauthoryear{{Zhu}, {Xu}, {You}, and {Bai}}{{Zhu}
  et~al\mbox{.}}{2020}]%
        {groupdnet}
\bibfield{author}{\bibinfo{person}{Zhen {Zhu}}, \bibinfo{person}{Zhiliang
  {Xu}}, \bibinfo{person}{Ansheng {You}}, {and} \bibinfo{person}{Xiang {Bai}}.}
  \bibinfo{year}{2020}\natexlab{}.
\newblock \showarticletitle{Semantically Multi-Modal Image Synthesis}. In
  \bibinfo{booktitle}{\emph{2020 IEEE/CVF Conference on Computer Vision and
  Pattern Recognition (CVPR)}}. \bibinfo{pages}{5467--5476}.
\newblock


\end{thebibliography}
